\documentclass[aps,prd,twocolumn,groupedaddress,amssymb,eqsecnum,showpacs,
nofootinbib]{revtex4}
\usepackage{graphicx}
\usepackage{mathptmx}
\usepackage{bm}
\usepackage{times}

\begin{document}

\preprint{UMHEP-462,CMU-HEP-06-07}

\title{The renormalization of the energy-momentum tensor for an effective initial state}

\author{Hael Collins}
\email{hael@physics.umass.edu}
\affiliation{Department of Physics, University of Massachusetts, 
Amherst MA\ \ 01003}
\author{R.~Holman}
\email{rh4a@andrew.cmu.edu}
\affiliation{Department of Physics, Carnegie Mellon University, 
Pittsburgh PA\ \ 15213}

\date{\today}

\begin{abstract}
An effective description of an initial state is a method for representing the signatures of new physics in the short-distance structure of a quantum state.  The expectation value of the energy-momentum tensor for a field in such a state contains new divergences that arise when summing over this new structure.  These divergences occur only at the initial time at which the state is defined and therefore can be cancelled by including a set of purely geometric counterterms that also are confined to this initial surface.  We describe this gravitational renormalization of the divergences in the energy-momentum tensor for a free scalar field in an isotropically expanding inflationary background.  We also show that the backreaction from these new short-distance features of the state is small when compared with the leading vacuum energy contained in the field.
\end{abstract}


\maketitle

\section{Introduction}
\label{intro}

It is fairly typical for a quantum field theory to produce infinities in the process of extracting its classical predictions.  The most familiar example of this property is the set of divergences that appear in the perturbative evaluation of a matrix element in an interacting field theory.  Yet divergences also occur in a perfectly free field theory, when we consider its effect on the classical background through which it propagates.  In flat space, the sum of the contributions to the ground state energy density is infinite and further infinities, accompanying geometric invariants which would have vanished in flat space, arise in a curved background.

In both cases, the space-time or field theory structures that accompany the divergences precisely match the structures already present in other parts of the theory.  In the former case they match the effects produced by local operators contained in the Lagrangian of the field theory and in the latter they match the geometric invariants that compose the gravitational action.  So by appropriately rescaling the parameters of the theory, the masses and couplings of the quantum field theory or the coefficients of the terms in the gravitational action, we can completely absorb all of these divergences and what remains is a renormalized theory whose predictions are finite.  In the process we lose the idea of a fixed constants parameterizing the theory, and which apply to all scales, and instead we discover that these parameters acquire a dependence on the scale at which we have defined the theory.

In choosing a particular state for extracting the classical predictions from a quantum theory, we are always making an implicit assumption about what happens at all scales, both those at which we know empirically that our theory provides a good description of nature as well as those at much shorter scales than have been probed.  This ignorance is usually permissible since the dynamics that are important at different scales often largely decouple from each another---the details and behavior of a theory at very short distance scales have a small effect on the measurements made at the larger scales.  A knowledge of the properties relevant at short distances thus only becomes necessary once we are able to make our long-distance measurements to a sufficient precision or if the dynamics of the system we are examining, to some measure, itself weakens this decoupling.  This principle of decoupling is implemented rigorously in this instance by constructing an effective theory description of the state of a system.

The basic philosophy of the effective state idea is that we ought to include a general set of short-distance structures in a state to parameterize the possible differences between the state we have chosen, derived from what we know about nature down to some length scale, with the true state of the system.  Processes that sum over all the scales in the state---such as the intermediate virtual momenta in a loop correction or such as occur when adding together all of the energy and momentum contained within a field configuration---can produce new divergences that come when we sum over these short-distance structures.  A complete effective treatment then should show how to renormalize these divergences, yielding in the end a finite perturbative description, and also how to estimate the natural size any new features of the state would have in a particular measurement.

Although this effective treatment \cite{greens,dSgreens,schalm,schalm2,ekp,emil} of a state is newer and less familiar than the standard formulation of effective field theory \cite{eft}, together they capture different sides of what should be regarded as a single, complete effective description of a process.  A prediction based on the expectation value of a quantum operator requires knowing both the state of the system and how this state and operator evolve.  The standard treatment of effective field theory applies specifically to this {\it evolution\/}; it is implemented by including all the operators consistent with the degrees of freedom available and their symmetries.  The main difference is that in some settings the short-distance features of the state can have the larger effect on a long-distance measurement.  More precisely, for a measurement being made at an energy $E$ meant to see new dynamics at a mass scale $M$, whereas the corrections to the evolution tend to be suppressed by powers of $E^2/M^2$ \cite{kaloper}, the corrections to the state are often only suppressed by powers of $E/M$ \cite{cliff,brandenberger,gary,transplanck,fate}.  

Each of these two components of an effective theory has particular short-distance effects for which it is better suited to describe.  As a simple example \cite{cliff}, an effective state is particularly applicable for a light field that is coupled to a heavy field in an excited state.  Even when the heavy field cannot be directly observed, the state of the lighter field inherits some signature of the excited state to which it is coupled.  Moreover, since we shall define an effective state in terms of its eigenmodes, it is also ideally suited for studying the signatures of theories \cite{gary,brandenberger,transplanck,fate} that propose that some symmetry of nature is broken, deformed, or replaced by another principle at a fundamental energy scale, $M$.  Each of these examples---which require making some explicit assumption about the physics at this scale---do in fact produce corrections to long distance measurements that are only suppressed by $E/M$.  Compared with these specific cases, the virtue of an effective theory is that we can simultaneously explore all these possibilities, as well as others not previously contemplated, without restricting to a particular choice, if we start from a sufficiently general description of the structure of the state at tiny length scales.

The most important application of the effective state idea is to inflation \cite{textbooks} and its trans-Planckian problem \cite{brandenberger}.  Looking backwards in time, up to a certain point the subsequent growth of the structures in the universe and the pattern of anisotropies in the cosmic background radiation can be explained well by physics which has been empirically established in other settings \cite{dodelson}.  Prior to this point, the only input we need is that there exists some initial spectrum of primordial perturbations, extending well beyond the horizon at that time, in an otherwise very smooth, nearly isotropic universe.  The standard explanation for generating these primordial perturbations is through inflation, a superluminal phase of expansion sustained by the vacuum energy of one or more quantum fields.  As in any quantum field theory, there is an inherent variability in the fields and these quantum fluctuations provide the origin for the primordial perturbations as they first occur at some early stage of inflation, are stretched outside the horizon to be essentially frozen into the background, and re-enter the horizon again only after inflation has ended.

The expansion during inflation, which provides this elegant explanation for the original fluctuations in the background, has a rather peculiar consequence if we continue to look yet further back in time.  To fit all of the observed universe within a single causally connected region at some point during inflation typically requires that, subsequent to that time, the universe should expand by at least a factor of about $10^{26}$ to $10^{30}$ before inflation comes to an end.  Most inflationary models have no difficulty producing this much expansion and usually produce substantially more.  Yet if inflation does last a bit longer, then following backwards in time one of the quantum fluctuations responsible for a primordial perturbation that led to structures being observed in the universe today, it would have inevitably had its origin at a scale smaller than the Planck length.  This observation---the seeming dependence of the primordial perturbation spectrum on physics beyond the Planck scale---constitutes the trans-Planckian problem of inflation \cite{brandenberger}.  

The effective-state approach addresses this problem by not attempting to follow inflation arbitrarily far back but instead by choosing a general description of the state of the quantum fields at a sufficiently early time during inflation.  In practice, the time at which we define the state should be early enough that the features we wish to predict are within the horizon at that time but not so early that they are smaller than the Planck length.  We shall frequently describe the state defined thus to be an {\it effective initial state\/}, but we are actually referring to the earliest time at which---in the absence of an empirically verified fundamental theory for gravity---the theory is predictive and not to the beginning of inflation.  The signatures of earlier times can be encoded in the structure of the state but it is important to note that the field theory never evaluates any quantity earlier than this ``initial time'' \cite{schwinger}, unlike what would be done in an $S$-matrix calculation.

This article examines how the presence of the short-distance structures of an effective state of a quantum scalar field affect the renormalization of the classical gravitational action.  This renormalization is done both to show the overall consistency of this effective theory and also to provide a comparison between the size of the renormalized contributions from this structure and the vacuum energy that sustains inflation.  We discover that the additional features in the state, using the standard vacuum as a reference state, produce divergences in the energy-momentum tensor of the field which are proportional to purely gravitational objects and which are confined to the initial-time hypersurface where the effective state is defined.  These divergences are thus removed by including initial-boundary counterterms in the gravitational action.  Once done, the renormalized gravitational equations of motion are thereafter finite.  

We begin in Sec.~\ref{effectivestate} with a brief review of the effective state idea and its construction; the standard vacuum serves a role similar to that of the renormalizable operators in the usual formulation of an effective field theory, so we define a general effective state relative to this state.  We next evaluate the energy-momentum tensor, which provides the source for gravity, in a general effective state for an isotropically expanding background.  As is typical for an effective state, the divergences in the expectation value of the energy-momentum tensor fall into two classes, those that appear throughout the space-time and those that are confined to the initial time.  The former occur even for the standard vacuum state and require the renormalization of the bulk properties of gravity, which we show in detail in Sec.~\ref{bulk} as a preparation for treating the boundary divergences.

Because the short-distance structures generically break some of the space-time symmetries at those scales, the new gravitational counterterms only need to respect that subset of the symmetries that is left unbroken by the state.  The generic gravitational counterterms correspond to geometric invariants confined specifically to the initial-time boundary.  Section~\ref{boundary} illustrates this boundary renormalization first for the simplest case of an effective state in Minkowski space, where a simple surface tension term is sufficient to remove all the new state-dependent divergences in the energy and momentum of the field, before treating a general isotropically expanding background.  In this section we show how the renormalization proceeds explicitly for a few simple examples and then cover the higher order divergences from the state more generally.  There is a natural relation between the moments used to describe the structure of an effective state and the order, in powers of derivatives, of the gravitational invariants needed to renormalize the new boundary divergences.  We describe this power counting and estimate the generic backreaction from the short-distance structure of the state.  If the expansion rate during inflation is denoted by $H$, and $M_{\rm Pl}$ and $M$ correspond respectively to the Planck mass and to the scale of new physics responsible for that structure, we find that the size of this backreaction, relative to the vacuum energy sustaining the inflation, is suppressed at least by 
\begin{equation}
{H^2\over M_{\rm pl}^2} {H\over M} . 
\label{backreactint}
\end{equation} 
Section~\ref{conclude} ends with a few comments about future work \cite{next} and about how the renormalization proceeds if we also treat gravity as a quantum theory.  In a tree calculation of this latter approach, the gravitational boundary counterterms can be integrated more readily than in the more standard classical approach described here.

\section{An effective description of an initial state}
\label{effectivestate}

Let us consider a free scalar field $\varphi(x)$ that propagates in a curved background.  The most general action that is quadratic in the field and contains no more than two derivatives has the form, 
\begin{equation}
S_\varphi = \int d^4x\, \sqrt{-g} \bigl[ 
{\textstyle{1\over 2}} g^{\mu\nu} \partial_\mu\varphi\partial_\nu\varphi 
- {\textstyle{1\over 2}} \xi R \varphi^2 
- {\textstyle{1\over 2}} m^2 \varphi^2 \bigr] . 
\label{action}
\end{equation}
The coupling between the curvature $R$ and the field does not usually have a significant role in inflation and moreover complicates the structure of the equations of motion of the theory.  Therefore, we shall examine the simplest case of a minimally coupled scalar field, which corresponds to $\xi=0$.

Varying the minimally coupled action with respect to the field $\varphi$ produces the Klein-Gordon equation, 
\begin{equation}
\bigl[ \nabla^2 + m^2 \bigr] \varphi = 0 .
\label{KGeqn}
\end{equation}
The background during inflation is assumed to be spatially isotropic, and empirical evidence \cite{wmap} suggests that it is spatially flat as well, with any spatial variation treated as a small perturbation.  Through an appropriate definition of the time coordinate, such a space-time can be described quite generally by a conformally flat Robertson-Walker metric, 
\begin{equation}
ds^2 = g_{\mu\nu}\, dx^\mu dx^\nu 
= a^2(\eta)\, \bigl[ d\eta^2 - d\vec x\cdot d\vec x \bigr] , 
\label{metric}
\end{equation}
where the time-dependent expansion is described by the evolution of the scale factor $a(\eta)$. 

Because the metric is spatially flat, the spatial eigenmodes of the Klein-Gordon operator can be expressed as plane waves, which allows us to write the operator expansion for the field $\varphi$ in terms of its Fourier transform as 
\begin{equation}
\varphi(\eta,\vec x) = \int {d^3\vec k\over (2\pi)^3}\, 
\bigl[ \varphi_k(\eta) e^{i\vec k\cdot\vec x} a_{\vec k} 
+ \varphi_k^*(\eta) e^{-i\vec k\cdot\vec x} a_{\vec k}^\dagger \bigr] , 
\label{opexpand}
\end{equation} 
where the mode functions $\varphi_k(\eta)$ are the solutions to the equation 
\begin{equation}
\varphi_k^{\prime\prime} + 2aH \varphi_k^{\prime} 
+ \bigl( k^2 + a^2m^2 \bigr) \varphi_k = 0 . 
\label{KGmodes}
\end{equation}
Here we have used a prime to denote a derivative with respect to the conformal time $\eta$ and $H$---the Hubble scale---represents the fractional rate of change of the scale factor $a(\eta)$, 
\begin{equation}
H = {a'\over a^2} . 
\label{hubble}
\end{equation}
The equal time commutation relation between the field and its canonically conjugate momentum essentially fixes the normalization of the modes so that a general solution to the Klein-Gordon equation for the modes, subject to this normalization constraint, is completely specified by a single constant of integration, $f_k$, 
\begin{equation}
\varphi_k(\eta) = {1\over\sqrt{1-f_kf_k^*}} \bigl[ 
U_k(\eta) + f_k U_k^*(\eta) \bigr] . 
\label{genmodes}
\end{equation}
We have implicitly written this expression for the modes of a general state in terms of those of a particular state.  In the effective theory description of the {\it evolution\/} of a theory, the signals of the more fundamental theory are represented as nonrenormalizable corrections to a particular renormalizable Lagrangian, determined by the known fields and their interactions.  Similarly, in the effective description of a state, the signals of new physics are added as corrections, whose significance grows at shorter length scales, to a particular state.  The most natural choice for this state is the standard vacuum \cite{bunch}, whose mode functions we have denoted by $U_k(\eta)$.  The extra structure, represented by $f_k$, then encodes how the state departs from this vacuum at distances shorter than those that have been experimentally probed.  

Since we have defined a general state in terms of the standard vacuum, we should be a little more explicit as to the form of the vacuum mode functions, $U_k(\eta)$.  The general prescription that determines these modes is to select the maximally symmetric solution to the Klein-Gordon equation that matches the Poincar\'{e} invariant vacuum of Minkowski space at distances where the curvature of the background is not much apparent, $k\gg H$.  To represent this state more clearly, we introduce a real, generalized frequency function $\Omega_k(\eta)$ that determines the modes through
\begin{equation}
U_k(\eta) = {e^{-i\int_{\eta_0}^\eta d\eta'\, \Omega_k(\eta')}
\over a(\eta) \sqrt{2\Omega_k(\eta)}} . 
\label{adiabat}
\end{equation}
Superficially, this expression already begins to resemble the structure of the Minkowski vacuum modes, except that it is complicated by the time-dependence of the background.  The Klein-Gordon equation of Eq.~(\ref{KGmodes}) can be equivalently be written as a differential equation for $\Omega_k(\eta)$,
\begin{equation}
\Omega_k^2 = k^2 + a^2m^2 - {a^{\prime\prime}\over a} 
- {1\over 2} {\Omega_k^{\prime\prime}\over\Omega_k} 
+ {3\over 4} \left( {\Omega_k^\prime\over\Omega_k} \right)^2 . 
\label{KGOmega}
\end{equation}
Assuming for the moment, as will soon be justified, that the derivatives of $\Omega_k(\eta)$ are small as $k\to\infty$, in this limit 
\begin{equation}
\Omega_k \to k 
\qquad\hbox{as $k\to\infty$,} 
\label{Minklimit}
\end{equation}
so the modes do indeed become proportional to the standard Minkowski modes, 
\begin{equation}
U_k(\eta) \to {1\over a(\eta)} {e^{-ik(\eta-\eta_0)}\over\sqrt{2k}} 
\qquad\hbox{as $k\to\infty$.} 
\label{Minkmodes}
\end{equation}

Solving for the form of the modes, whether in the form $U_k(\eta)$ or through $\Omega_k$, can be quite difficult in general.  However, here we wish to examine the role of the short-distance behavior of the energy-momentum contained in the field $\varphi$ in the renormalization of the gravitational component of the theory.  Therefore, it is sufficient to have a solution of the modes only in the limit of large momenta, $k\to\infty$, so we introduce an approximation scheme that captures this behavior perturbatively.

We shall apply an adiabatic expansion, which assumes that time derivatives are small compared to the scales of interest.  Higher order terms are thus generated by taking the time derivatives of the lower order terms.  For example, if we denote the terms of this expansion by 
\begin{equation}
\Omega_k(\eta) = \Omega_k^{(0)}(\eta) + \Omega_k^{(1)}(\eta) 
+ \Omega_k^{(2)}(\eta) + \cdots , 
\label{OmegaNdef}
\end{equation}
where the leading term is 
\begin{equation}
{\Omega_k^{(0)}}^2 = \omega_k^2 - {a^{\prime\prime}\over a} , 
\label{Omega0def}
\end{equation}
with 
\begin{equation}
\omega_k^2 = k^2 + a^2m^2 , 
\label{omegadef}
\end{equation}
then the next order term is then determined by 
\begin{equation}
\Omega_k^{(1)} = 
- {1\over 4} {\Omega_k^{(0)\prime\prime}\over\Omega_k^{(0) 2}} 
+ {3\over 8} {1\over \Omega_k^{(0)}} 
\biggl( {\Omega_k^{(0)\prime}\over\Omega_k^{(0)}} \biggr)^2 . 
\label{Omega1def}
\end{equation}
Although the mass is usually small in inflation, we have kept it in our definition of $\omega_k$ since it naturally regulates the behavior of the momentum integrals at long wavelengths and thereby prevents any infrared divergences.  Here, we shall not usually require any terms which decay as $k^{-5}$ or more rapidly; thus, 
\begin{eqnarray}
\Omega_k^{(1)} 
&\!\!\!=\!\!\!& - {1\over 4} {a^2m^2\over\omega_k^3} \biggl[ 
{a^{\prime\prime}\over a} - \biggl( {a'\over a} \biggr) \biggr]
\nonumber \\
&&
+\ {1\over 8} {1\over\omega_k^3}
\biggl\{ {a^{\prime\prime\prime\prime}\over a} 
- 2 {a^{\prime\prime\prime}\over a} {a'\over a} 
- \biggl( {a^{\prime\prime}\over a} \biggr)^2 
+ 2 {a^{\prime\prime}\over a} \biggl( {a'\over a} \biggr)^2 
\biggr\}  
\nonumber \\
&&
+\ {\cal O} \biggl( {1\over k^5} \biggr) , 
\label{Omega1approx}
\end{eqnarray}
for example.

We also require a more explicit way of representing the structure of the initial state.  Because the standard vacuum provides such a good fit with the observed fluctuations in the microwave background and the large scale structure, $f_k$ should vanish at long wavelengths, $f_k\to 0$ as $k\to 0$.  However, in the opposite limit, new effects can distort the form of the state we have chosen for our vacuum, since such a choice was implicitly based on assumptions about the interactions, the behavior of gravity and the symmetries of the theory at scales well beyond those that have been experimentally tested.  The additional structure we have included through the function $f_k$ is meant to describe the possible difference between the true vacuum state and the standard vacuum.  The scales often assumed for the Hubble scale $H$ during inflation can be as high as just a few orders of magnitude below the Planck scale, $M_{\rm pl}$.  Thus the standard vacuum is chosen by appealing to the behavior of the field in a regime approaching that of the Planck scale, where the role and dynamics of gravity are not understood.  Even for inflation with a much lower value of $H$, the state might still contain new structure, $f_k\not= 0$, through its excitations or its interactions with other hidden fields.

We therefore introduce a scale $M$ at which the new features of the state appear.  Most conservatively, we might have $M=M_{\rm pl}$; but most generally, depending upon what generates this apparent new structure, $M$ could even be significantly below $M_{\rm pl}$.  As a simple example, we shall most often use a representation of $f_k$ which does does not break the underlying spatial isotropy of the theory and which vanishes at long wavelengths, such as 
\begin{equation}
f_k = \sum_{n=1}^\infty d_n {k^n\over (aM)^n} , 
\label{fkdef}
\end{equation}
although to illustrate how the divergences associated with a general state are renormalized, it will sometimes also be more transparent to use a state whose structure is modified at long wavelengths.  Such states too can have divergences in the energy-momentum tensor since, just as in the case of the bulk renormalization which has quartic divergences even for a free scalar field theory, the divergences encountered in renormalizing this tensor are usually more severe than those we encounter in the more standard setting of renormalizing an interacting field theory. 

To specify the structure in the state, we establish an initial condition on the modes, 
\begin{equation}
n^\mu\nabla_\mu \varphi_k(\eta) \bigr|_{\eta=\eta_0} 
= -i \varpi_k \varphi_k(\eta_0) , 
\label{initial}
\end{equation}
where $n_\mu$ is a time-like unit normal vector defined along the surface $\eta=\eta_0$.  For the spatially isotropic metric in Eq.~(\ref{metric}), \begin{equation}
n_\mu = \bigl( a(\eta), 0, 0, 0 \bigr) . 
\label{ndef}
\end{equation}
To obtain the simple structure for a general mode function given before, in Eq.~(\ref{genmodes}), the function $\varpi_k$ is related to the structure of the state $f_k$ by
\begin{equation}
\varpi_k \equiv {1-f_k\over 1+f_k} {\Omega_k\over a} - i H - {i\over 2} {\Omega'_k\over\Omega_k} . 
\label{varpidef}
\end{equation}

The Green's function describing the propagation of information associated with a point source should also be consistent with the initial state and the initial condition.  If we denote the effective state by $|0_{\rm eff}\rangle$, then this propagator is the time-ordered expectation value of two fields
\begin{eqnarray}
\langle 0_{\rm eff} | T \bigl( \varphi(x) \varphi(x') \bigr) | 0_{\rm eff} \rangle
&\!\!\!=\!\!\!& -i G_F(x,x') 
\label{feynman} \\
&\!\!\!=\!\!\!& 
-i \int {d^3\vec k\over (2\pi)^3}\, e^{i\vec k\cdot(\vec x-\vec x')} 
G_k(\eta,\eta') , 
\nonumber 
\end{eqnarray}
where the time-ordering is generalized to satisfy the same initial condition as in Eq.~(\ref{initial}), 
\begin{eqnarray}
n^\mu(\eta)\nabla_\mu G_k(\eta,\eta') \bigr|_{\eta=\eta_0} 
&\!\!\!=\!\!\!& i \varpi_k^* G_k(\eta_0,\eta') 
\nonumber \\
n^\mu(\eta')\nabla^\prime_\mu G_k(\eta,\eta') \bigr|_{\eta'=\eta_0} 
&\!\!\!=\!\!\!& i \varpi_k^* G_k(\eta,\eta_0) . 
\label{intiprop}
\end{eqnarray}
These conditions together introduce additional structure in the propagator, which takes the form 
\begin{eqnarray}
-i G_k(\eta,\eta') 
&\!\!\!=\!\!\!& \Theta(\eta-\eta')\, U_k(\eta)U_k^*(\eta') 
\nonumber \\
&&
+ \Theta(\eta'-\eta)\, U_k^*(\eta)U_k(\eta') 
\nonumber \\
&&
+ f_k^* U_k(\eta)U_k(\eta') . 
\label{propagator} 
\end{eqnarray}
Note that the propagator is still the Green's function for a point source; however, it does contain additional terms which represent the forward propagation of the structure which we have defined at the initial time.  This propagator is the basis for our effective theory description of the state and is implicitly used whenever we take the expectation value of a product of scalar fields, such as that of the energy-momentum tensor.  

A complete derivation of the propagator for this effective theory is contained in \cite{greens,dSgreens}, where both the consistency of the time-ordering with the initial condition at $\eta=\eta_0$ and the physical meaning of the terms in the propagator are explained in detail.  To provide a basis for a controlled perturbation theory, it is essentially unique, as other attempts to include state-dependent effects in the propagator, such as those examined \cite{fate,einhorn} for the invariant $\alpha$-states of de Sitter space \cite{alpha}, lead to various related pathologies.

\section{The renormalization of the gravitational action}
\label{bulk}

When a quantum field theory is placed in a gravitational background, even a strictly classical one, the short-distance divergences associated with the field require a rescaling of the gravitational theory \cite{books,adiabatic}.  The most familiar example of this phenomenon is the vacuum energy associated with the field.  This energy density diverges in flat space, but it is usually neglected by adding a bare cosmological constant to the theory to cancel precisely the contribution that came by summing over the infinite set of modes of the field, or equivalently by normally ordering the operators when evaluating a matrix element. 

In a curved background, over distances where the curvature changes little, the space-time looks locally flat and so the same redefinition of the cosmological constant is required except that the net result can now be nonzero.  The energy and momentum contained within the field also produce other divergences which are proportional to how the background changes.  A similar procedure as that applied to remove the vacuum energy density divergence is used to treat these divergences; in the process, the parameters of the gravitational theory are redefined so that the sum of these gravitational parameters and the corresponding divergences that arose in summing over the infinitesimally short-distance modes of the quantum field yields a finite set of gravitational equations of motion.  If the field and, for the treatment here, the state preserve the background symmetries of the theory, all of the divergences from the field will be exactly proportional to some curvature invariant.  However, when the state breaks some of these symmetries, new terms in the gravitational action will be required to remove the short-distance divergences associated with this state.

Since a general initial state is defined relative to the standard vacuum, we first show how the renormalization of a field in this nominal vacuum state proceeds in a simple curved background, partially to illustrate our renormalization procedure in a familiar setting but also to show that the renormalization of the bulk parameters of the gravitational action is unaffected by the short-distance details of the effective state.

As an effective theory, the gravitational action, invariant under general changes of coordinates, has the form,\footnote{Here we specify the sign conventions that we use.  The Riemann curvature tensor is defined by $-R^\lambda_{\ \mu\nu\rho} = \partial_\rho \Gamma^\lambda_{\mu\nu} - \partial_\nu \Gamma^\lambda_{\mu\rho} +  \Gamma^\lambda_{\rho\sigma} \Gamma^\sigma{\mu\nu} - \Gamma^\lambda_{\nu\sigma} \Gamma^\sigma_{\mu\rho}$ and the Ricci tensor is given by $R^\lambda_{\ \mu\lambda\nu} = R_{\mu\nu}$.} 
\begin{eqnarray}
S_g &\!\!\!=\!\!\!& \int d^4x\, \sqrt{-g}\, \bigl[ 
2\Lambda + M_{\rm pl}^2 R + \alpha R^2 
\label{gravactionfull} \\
&&\qquad\qquad
+\ \beta R_{\mu\nu}R^{\mu\nu} 
+ \gamma R_{\lambda\mu\nu\rho}R^{\lambda\mu\nu\rho}
+ \cdots
\bigr] . 
\nonumber 
\end{eqnarray}
The Planck mass, $M_{\rm pl}$, can be equivalently given in terms of Newton's constant, $G$, 
\begin{equation}
M_{\rm pl}^2 \equiv {1\over 16\pi G} . 
\label{Mpldef}
\end{equation}
The gravitational action is arranged as series of terms that contain successively more derivatives of the metric; if the metric changes slowly over lengths of the order $M_{\rm pl}^{-1}$, then terms with more derivatives are naturally suppressed.  To treat the divergences in the energy-momentum tensor of a free scalar field, only terms that contain four or fewer derivatives of the metric are sufficient.  Two linear combinations of the fourth-order terms have an additional geometric significance.  One of these, the Gauss-Bonnet term, 
\begin{equation}
R^2 - 4 R_{\mu\nu}R^{\mu\nu} 
+ R_{\lambda\mu\nu\rho}R^{\lambda\mu\nu\rho} , 
\label{gaussbonnet}
\end{equation}
is a topological invariant in four dimensions while the other, constructed entirely from the Weyl tensor,
\begin{equation}
C_{\lambda\mu\nu\rho}C^{\lambda\mu\nu\rho} 
= {\textstyle{1\over 3}} R^2 - 2 R_{\mu\nu}R^{\mu\nu} 
+ R_{\lambda\mu\nu\rho}R^{\lambda\mu\nu\rho} , 
\label{weylweyl}
\end{equation}
vanishes for conformally flat metrics such as that of a Robertson-Walker space-time, which we have explicitly written in a conformally flat form in Eq.~(\ref{metric}).  

As a consequence, to treat the divergences in the energy-momentum tensor in an isotropically expanding space-time only a single linear combination of the four-derivative terms, orthogonal to both the Gauss-Bonnet and the Weyl terms, is required.  The simplest choice is to use just the $R^2$ term alone, 
\begin{equation}
S_g = \int d^4x\, \sqrt{-g}\, \Bigl[ 
2 \Lambda + M_{\rm pl}^2 R + \alpha R^2 + \cdots
\Bigr] . 
\qquad
\label{gravaction}
\end{equation}
Varying this component of the action with respect to a small change in the metric yields, 
\begin{eqnarray}
&&\!\!\!\!\!\!\!\!\!\!\!\!\!
{2\over\sqrt{-g}} {\delta S_g\over\delta g^{\mu\nu}} 
\label{gravityEOM} \\
&\!\!\!=\!\!\!& 2\Lambda g_{\mu\nu} 
- 2M_{\rm pl}^2 \bigl[ 
R_{\mu\nu} - {\textstyle{1\over 2}} g_{\mu\nu} R \bigr] 
\nonumber \\
&& 
-\ 4\alpha \bigl[
\nabla_\mu\nabla_\nu R - g_{\mu\nu} \nabla^2 R 
+ RR_{\mu\nu} - {\textstyle{1\over 4}} g_{\mu\nu} R^2 
\bigr] . 
\nonumber 
\end{eqnarray}

The scalar field $\varphi(x)$ introduced in the last section provides the source for the curvature of the background.  Varying its contribution to the action in Eq.~(\ref{action}) yields the energy-momentum tensor,
\begin{equation}
T_{\mu\nu} = - {2\over\sqrt{-g}} {\delta S_\varphi\over\delta g^{\mu\nu}} , 
\label{curveEMT}
\end{equation}
which is 
\begin{equation}
T_{\mu\nu} = \partial_\mu\varphi \partial_\nu\varphi 
- {\textstyle{1\over 2}} g_{\mu\nu} 
   g^{\lambda\rho} \partial_\lambda\varphi \partial_\rho\varphi 
+ {\textstyle{1\over 2}} g_{\mu\nu} m^2 \varphi^2 . 
\label{EMT}
\end{equation}
This tensor, unlike that obtained by the variation of the purely classical gravitational action, is an operator; so to use it consistently in the gravitational equations of motion, we must first evaluate its expectation value in the state $|0_{\rm eff}\rangle$, 
\begin{eqnarray}
&&\!\!\!\!\!\!\!\!\!\!\!\!\!
\langle 0_{\rm eff} | T_{\mu\nu} | 0_{\rm eff} \rangle 
\label{fullEOM} \\
&\!\!\!=\!\!\!&
2 g_{\mu\nu} \Lambda 
-2 M_{\rm pl}^2 \bigl[ R_{\mu\nu} - {\textstyle{1\over 2}} g_{\mu\nu} R \bigr] 
\nonumber \\
&&
-\ 4\alpha \bigl[ 
\nabla_\mu\nabla_\mu R - g_{\mu\nu} \nabla^2 R + R R_{\mu\nu} 
- {\textstyle{1\over 4}} g_{\mu\nu} R^2 \bigr] 
\nonumber 
\end{eqnarray}
In a spatially flat background, the field has the same symmetries; this property implies that the expectation value of the energy-momentum tensor is diagonal, with components independent of the spatial position, 
\begin{equation}
\langle 0_{\rm eff} | T_\mu^{\ \nu}(x) | 0_{\rm eff} \rangle \equiv 
{\rm diag}\bigl[ \rho(\eta), -p(\eta), -p(\eta), -p(\eta) \bigr]
\label{Texptdef}
\end{equation}
The functions $\rho(\eta)$ and $p(\eta)$ represent the energy density and the pressure associated with the field $\varphi$ and they are related through the conservation equation,
\begin{equation}
\nabla_\lambda T^\lambda_{\ \mu} = 0 , 
\label{conserveEMT}
\end{equation}
which imposes only a single relation between the pressure and density, 
\begin{equation}
\rho' = - 3aH(\rho+p) .
\label{conserved}
\end{equation}
Note that the pressure is thereby completely determined by the density.

When we evaluate the gravitational equations of motion in Eq.~(\ref{fullEOM}) for an isotropically expanding space-time, as given in Eq.~(\ref{metric}), we obtain two equations:  one for the purely temporal component,
\begin{eqnarray}
\rho 
&\!\!\!=\!\!\!& 
2 \Lambda + M_{\rm pl}^2 {6\over a^2} \biggl( {a'\over a} \biggr)^2 
\nonumber \\
&&
+ {36\alpha\over a^4} \biggl\{ 
2 {a^{\prime\prime\prime}\over a} {a'\over a} 
- \biggl( {a^{\prime\prime}\over a} \biggr)^2 
- 4 {a^{\prime\prime}\over a} \biggl( {a'\over a} \biggr)^2 
\biggr\} \qquad
\label{00EOM}
\end{eqnarray}
and a second for the purely spatial components of the equation,
\begin{eqnarray}
p &\!\!\!=\!\!\!& - 2\Lambda + M_{\rm pl}^2 {2\over a^2} \biggl[ 
\biggl( {a'\over a} \biggr)^2 - 2 {a^{\prime\prime}\over a} 
\biggr]
\label{ijEOM} \\
&&
- {24\alpha\over a^4} 
\biggl\{ {a^{\prime\prime\prime\prime}\over a} 
- 5 {a^{\prime\prime\prime}\over a} {a'\over a} 
- {5\over 2} \biggl( {a^{\prime\prime}\over a} \biggr)^2 
+ 8 {a^{\prime\prime}\over a} \biggl( {a'\over a} \biggr)^2 
\biggr\} . 
\nonumber 
\end{eqnarray}
Each of these equations contains three scale-factor-dependent structures which multiply the three parameters of the gravitational action:  the cosmological constant, which has a trivial prefactor, the Planck mass and $\alpha$, the coefficient of the $R^2$ term.  The divergences in the field-dependent side of these equations will reproduce each of these scale-dependent structures exactly, which means that they can be cancelled through an appropriate rescaling of the gravitational parameters.

The components of the energy-momentum tensor contain two general classes of contributions, those which are universal, which assume the same form regardless of the effective state, and those which depend upon our specific choice of this initial state.  The former are responsible for the familiar need to renormalize the standard parameters that describe the gravitational theory while the latter class only produces divergences at the initial time, where the state was defined, and so they essentially correspond to a renormalization of the state.  In terms of how we have constructed the propagator, given in Eq.~(\ref{propagator}), these two types of divergences can alternatively be described as those that arise from the part of the state that preserves Poincar\'{e} invariance at infinitesimal lengths and those that come from summing over the short-distance structures that depart from this invariance.  This observation is important since it means that in renormalizing the new boundary divergences, we should allow a larger class of operators among the boundary counterterms than is ordinarily considered since they need only to respect the symmetries of the state, and not that of the full theory.

To learn how structure in the effective initial state affects the expectation value of the energy-momentum tensor, we shall write it in terms of the propagator.  To do so, we first write the two fields in $T_{\mu\nu}$ temporarily at different points \cite{pointsplit}.  The idea behind writing the point-split version of the expectation value of $T_{\mu\nu}$ in terms of the propagator is that it already properly incorporates the effects of the initial state, represented by the final term of Eq.~(2.23).  Applying the generalized time-ordering that produced this propagator to other matrix elements of the theory---in this case, that of $T_{\mu\nu}$---thereby avoids non-renormalizable divergences similar to those that would appear in the loop corrections of an interacting scalar theory based on other propagator structures, such as those mentioned briefly in \cite{greens,dSgreens} and examined in more detail in \cite{fate,einhorn} in the setting of de Sitter space.  Thus, the essentially unique choice that yields a renormalizable expectation value for the energy-momentum tensor is precisely that where the state-dependent part is given by a derivative operator acting of the propagator of Eq.~(2.23), as in Eq.~(3.15) just below.  

The time-dependence of the state independent part is less constrained.  The only subtlety in choosing the propagator's time-ordering is that the energy-momentum tensor contains time derivatives, which act solely on the fields, so we are free to commute the derivatives with the operation of evaluating an expectation value as long as we do not allow them to act on the $\Theta$-functions associated with the time-ordering.  Thus, the contribution of the scalar field to the gravitational equations can be written as a derivative operator acting on the free propagator,
\begin{eqnarray}
&&\!\!\!\!\!\!
\langle 0_{\rm eff} | T_{\mu\nu}(x) | 0_{\rm eff} \rangle 
\nonumber \\
&&
= -i \lim_{x'\to x} \bigl[ 
\hat\partial_\mu \hat\partial'_\nu
- {\textstyle{1\over 2}} g_{\mu\nu} 
  g^{\lambda\rho} \hat\partial_\lambda \hat\partial'_\rho 
+ {\textstyle{1\over 2}} g_{\mu\nu} m^2 
\bigr] G_F(x,x') , 
\nonumber \\
&&
\label{EMTexp}
\end{eqnarray}
where $\hat\partial_\mu$ denotes a derivative that does not act on the $\Theta$-functions.  Using Eq.~(\ref{propagator}) for the initial state propagator, the energy density and pressure produced by the scalar field are given by 
\begin{eqnarray}
\rho(\eta) &\!\!\!=\!\!\!& 
{1\over 2} {1\over a^2} \int {d^3\vec k\over(2\pi)^3}\, 
\bigl\{ U'_k U^{*\prime}_k + (k^2+a^2m^2) U_k U^*_k 
\nonumber \\
&&\qquad\qquad
+ f_k^* \bigl[ U'_k U'_k + (k^2+a^2m^2) U_k U_k \bigr]
\bigr\} \qquad 
\label{rhofull}
\end{eqnarray}
and
\begin{eqnarray}
p(\eta) &\!\!\!=\!\!\!& - \rho(\eta) 
+ {1\over a^2} \int {d^3\vec k\over(2\pi)^3}\, 
\bigl\{ U'_k U^{*\prime}_k + {\textstyle{1\over 3}} k^2 U_k U^*_k 
\nonumber \\
&&\qquad\qquad\qquad\quad 
+ f_k^* \bigl[ U'_k U'_k + {\textstyle{1\over 3}} k^2 U_k U_k \bigr]
\bigr\} . \qquad 
\label{pfull}
\end{eqnarray}
Because of the conservation equation, Eq.~(\ref{conserved}), the pressure is already fixed by the energy density; but for completeness, we shall continue to include both.

This section illustrates how the short-distance divergences contained within $\rho$ and $p$ require the renormalization of the bulk properties of the gravitational theory.  For the remainder of this section we therefore restrict to the state-independent parts of the energy-momentum tensor, which are  
\begin{eqnarray}
\rho_{\rm bulk} 
&\!\!\!=\!\!\!& {1\over 2} {1\over a^2} \int {d^3\vec k\over(2\pi)^3}\, 
\Bigl[ U'_k U^{*\prime}_k + (k^2+a^2m^2) U_k U^*_k \Bigr] \qquad 
\nonumber \\
p_{\rm bulk} 
&\!\!\!=\!\!\!& - \rho_{\rm bulk} 
+ {1\over a^2} \int {d^3\vec k\over(2\pi)^3}\, 
\biggl[ U'_k U^{*\prime}_k + {1\over 3} k^2 U_k U^*_k \biggr] . 
\label{rhopbulk}
\end{eqnarray}

\begin{widetext}
We do not need the detailed form of the modes, but rather only their behavior at infinitesimally short distances where the size of the momentum $k$ is very large.  For this purpose, our adiabatic expansion of the previous section, Eq.~(\ref{adiabat}), is quite sufficient; applying this expansion, we discover three classes of divergences in the density,
\begin{eqnarray}
\rho_{\rm bulk}(\eta) 
&\!\!\!=\!\!\!& 
{1\over 2} {1\over a^4} \int {d^3\vec k\over (2\pi)^3}\, \omega_k 
+ {1\over 4} {1\over a^4} \biggl( {a'\over a} \biggr)^2 
\int {d^3\vec k\over (2\pi)^3}\, 
\biggl[ {1\over\omega_k} + {a^2 m^2\over\omega_k^3} \biggr] 
- {1\over 16} {1\over a^4} \biggl\{
2 {a^{\prime\prime\prime}\over a} {a'\over a}
- \biggl( {a^{\prime\prime}\over a} \biggr)^2
- 4 {a^{\prime\prime}\over a} \biggl( {a'\over a} \biggr)^2
\biggr\} 
\int {d^3\vec k\over (2\pi)^3}\, {1\over\omega_k^3} 
\nonumber \\
&&
+ {\rm finite}
\label{rhodiv}  
\end{eqnarray}
and pressure, 
\begin{eqnarray}
p_{\rm bulk}(\eta) 
&\!\!\!=\!\!\!& {1\over 6} {1\over a^4} \int {d^3\vec k\over (2\pi)^3}\, 
\biggl[ \omega_k - {a^2m^2\over\omega_k} \biggr]
- {1\over 12} {1\over a^4} \biggl[ 
\biggl( {a'\over a} \biggr)^2 - 2 {a^{\prime\prime}\over a}
\biggr] 
\int {d^3\vec k\over (2\pi)^3}\, {1\over\omega_k} 
\nonumber \\
&& 
+ {1\over 24} {1\over a^4} \biggl\{ {a^{\prime\prime\prime\prime}\over a} 
- 5 {a^{\prime\prime\prime}\over a} {a'\over a} 
- {5\over 2} \biggl( {a^{\prime\prime}\over a} \biggr)^2 
+ 8 {a^{\prime\prime}\over a} \biggl( {a'\over a} \biggr)^2 
\biggr\}
\int {d^3\vec k\over (2\pi)^3}\, {1\over\omega_k^3} 
+ {1\over 6} {1\over a^4} \biggl[ 
\biggl( {a'\over a} \biggr)^2 - {a^{\prime\prime}\over a} \biggr]  
\int {d^3\vec k\over (2\pi)^3}\, 
\left[ {2\over\omega_k} + {a^2 m^2\over\omega_k^3} \right]
\nonumber \\
&&
+ {\rm finite} . 
\label{pdiv}  
\end{eqnarray}
\end{widetext}
where, as before, 
\begin{equation}
\omega_k = \sqrt{k^2 + a^2(\eta) m^2} . 
\label{omegaagain}
\end{equation}
These divergences are quartic, quadratic and logarithmic in the momentum and respectively lead to the renormalization of the cosmological constant, the Planck mass and the coefficient of the $R^2$ term.  We have arranged the terms already so that their coefficients are similar to those of the gravitational terms in Eqs.~(\ref{00EOM}) and (\ref{ijEOM}).  The only term that does not seem to have the correct structure is the last one in the expression for the pressure; once we have regulated the momentum integrals, this term has a completely finite limit as we remove the regulator and so it actually has no effect on the renormalization of the theory.

All of the divergent momentum integrals have a similar general form,
\begin{equation}
I(0,n) = \int {d^3\vec k\over (2\pi)^3}\, {1\over\omega_k^n}
= \int {d^3\vec k\over (2\pi)^3}\, {1\over (k^2+a^2m^2)^{n/2}} , 
\label{drzero}
\end{equation}
which diverges when $n=3,1,-1,-3, \ldots$.  We regulate these integrals by analytically continuing the number of spatial dimensions to $3-2\epsilon$ which yields an integral that is finite,
\begin{eqnarray}
I(\epsilon,n) 
&\!\!\!=\!\!\!& \int {d^{3-2\epsilon}\vec k\over (2\pi)^{3-2\epsilon}}\, 
{(a\mu)^{2\epsilon}\over\omega_k^n}
\nonumber \\
&\!\!\!=\!\!\!& 
{\sqrt{\pi}\over 8\pi^2} 
{\Gamma(\epsilon - {3-n\over 2})\over \Gamma({n\over 2})}
\biggl[ {4\pi\mu^2\over m^2} \biggr]^\epsilon (am)^{3-n} , 
\label{dreps}
\end{eqnarray}
until we restore the $\epsilon\to 0$ limit.  A small but important point is that we have introduced a {\it comoving\/} renormalization scale $a\mu$ into the dimensionally continued version of the momentum integral, since the integral itself is over the comoving momenta.  This definition means that $\mu$ is the {\it physical\/} renormalization scale and one consequence of this structure is that appearances of the scale factor $a(\eta)$ cancel in the logarithms in the dimensionally regularized expressions for the energy density and the pressure.

Applying this regularization scheme to the divergences in the energy-momentum tensor, we find 
\begin{widetext}
\begin{eqnarray}
\rho_{\rm bulk}(\eta) 
&\!\!\!=\!\!\!& 
- {m^4\over 64\pi^2} \biggl[ {1\over\epsilon} + {3\over 2} - \gamma + \ln 4\pi + \ln{\mu^2\over m^2} \biggr]
+ {m^2\over 192\pi^2} {6\over a^2} \biggl( {a'\over a} \biggr)^2 
\biggl[ {1\over\epsilon} + 1 - \gamma + \ln 4\pi + \ln{\mu^2\over m^2} \biggr]
\nonumber \\
&& 
-\ {1\over 2304\pi^2} {36\over a^4} \biggl\{
2 {a^{\prime\prime\prime}\over a} {a'\over a}
- \biggl( {a^{\prime\prime}\over a} \biggr)^2
- 4 {a^{\prime\prime}\over a} \biggl( {a'\over a} \biggr)^2
\biggr\} 
\biggl[ {1\over\epsilon} - \gamma + \ln 4\pi + \ln{\mu^2\over m^2} \biggr]
+ {\rm finite}
\label{rhoeps}  
\end{eqnarray}
and
\begin{eqnarray}
p_{\rm bulk}(\eta) 
&\!\!\!=\!\!\!& {m^4\over 64\pi^2} 
\biggl[ {1\over\epsilon} + {3\over 2} - \gamma + \ln 4\pi + \ln{\mu^2\over m^2} \biggr] 
+ {m^2\over 192\pi^2} {2\over a^2} 
\biggl[ \biggl( {a'\over a} \biggr)^2 - 2 {a^{\prime\prime}\over a} \biggr] 
\biggl[ {1\over\epsilon} + 1 - \gamma + \ln 4\pi + \ln{\mu^2\over m^2} \biggr] 
\nonumber \\
&& 
+\ {1\over 2304\pi^2} {24\over a^4} 
\biggl\{ {a^{\prime\prime\prime\prime}\over a} 
- 5 {a^{\prime\prime\prime}\over a} {a'\over a} 
- {5\over 2} \biggl( {a^{\prime\prime}\over a} \biggr)^2 
+ 8 {a^{\prime\prime}\over a} \biggl( {a'\over a} \biggr)^2 
\biggr\}
\biggl[ {1\over\epsilon} - \gamma + \ln 4\pi + \ln{\mu^2\over m^2} \biggr]
+ {\rm finite} . 
\label{peps}  
\end{eqnarray}
\end{widetext}
The scale factor-dependent parts of each of these terms now match exactly with those of the gravitational contributions to the equations of motion, given in Eqs.~(\ref{00EOM}) and (\ref{ijEOM}).  Therefore the divergences that arose from the presence of the scalar field can be absorbed by new, renormalized gravitational parameters, 
\begin{eqnarray}
\Lambda_R(\mu) &\!\!\!=\!\!\!& \Lambda + {m^4\over 128\pi^2} 
\biggl[ {1\over\epsilon} + {3\over 2} - \gamma + \ln 4\pi + \ln{\mu^2\over m^2} \biggr] 
\nonumber \\
M_{\rm pl,R}^2(\mu) &\!\!\!=\!\!\!& M_{\rm pl}^2 - {m^2\over 192\pi^2} 
\biggl[ {1\over\epsilon} + 1 - \gamma + \ln 4\pi + \ln{\mu^2\over m^2} \biggr]
\nonumber \\
\alpha_R(\mu) &\!\!\!=\!\!\!& \alpha + {1\over 2304\pi^2} 
\biggl[ {1\over\epsilon} - \gamma + \ln 4\pi + \ln{\mu^2\over m^2} \biggr] . 
\label{Zdefs}
\end{eqnarray}
We have applied a $\overline{\rm MS}$ renormalization scheme where we remove some of the artifacts of the dimensional regularization procedure along with the poles.  Notice that the finite renormalized parameters have acquired a dependence on the renormalization scale $\mu$.

If we define the renormalized energy density and pressure, $\rho_R^{\rm bulk}(\eta)$ and $p_R^{\rm bulk}(\eta)$, to be these quantities with the divergent pieces---precisely those explicitly shown in Eqs.~(\ref{rhoeps}) and (\ref{peps})---subtracted, then we finally obtain a completely finite set of bulk gravitational equations of motion, 
\begin{eqnarray}
\rho_R^{\rm bulk} 
&\!\!\!=\!\!\!& 2 \Lambda_R + M_{\rm pl,R}^2 {6\over a^2} \biggl( {a'\over a} \biggr)^2 
\nonumber \\
&&
+ {36\alpha_R\over a^4} \biggl\{ 
2 {a^{\prime\prime\prime}\over a} {a'\over a} 
- \biggl( {a^{\prime\prime}\over a} \biggr)^2 
- 4 {a^{\prime\prime}\over a} \biggl( {a'\over a} \biggr)^2 
\biggr\} 
\nonumber \\
&&
\label{00EOMR}
\end{eqnarray}
and 
\begin{eqnarray}
&&\!\!\!\!\!\!\!\!\!\!
p_R^{\rm bulk} 
\nonumber \\
&\!\!\!=\!\!\!& - 2\Lambda_R + M_{\rm pl,R}^2 {2\over a^2} \biggl[ 
\biggl( {a'\over a} \biggr)^2 - 2 {a^{\prime\prime}\over a} 
\biggr]
\nonumber \\
&&
- {24\alpha_R\over a^4} 
\biggl\{ {a^{\prime\prime\prime\prime}\over a} 
- 5 {a^{\prime\prime\prime}\over a} {a'\over a} 
- {5\over 2} \biggl( {a^{\prime\prime}\over a} \biggr)^2 
+ 8 {a^{\prime\prime}\over a} \biggl( {a'\over a} \biggr)^2 
\biggr\} . 
\nonumber \\
&&
\label{ijEOMR} 
\end{eqnarray}

What this derivation has provided is a prescription for renormalizing the divergences in the matter component of the theory through a redefinition of the parameters of the gravitational theory.  In this procedure, we first isolated the various divergent parts of $\rho$, or equivalently $p$, by applying our adiabatic expansion of the modes.  The dependence on the scale factor of the prefactors of each class of divergences then determine which of the gravitational terms must be renormalized.  We next apply this same method to the renormalization of the divergences produced by summing over the short-distance structure of the state.  Since a general state might break some of the background symmetries, the class of allowed counterterms will be larger---for example including invariants constructed from the extrinsic curvature---and will therefore depend upon the specific state we have chosen.

\section{Boundary renormalization}
\label{boundary}

An effective initial state describes the possible differences between the nominal, standard vacuum state, which was the basis of the bulk renormalization just calculated, and the true state of the universe during inflation.  Depending upon what symmetries or principles hold at very short distances, this state, even if it corresponds to the true vacuum state of a fundamental theory, could differ substantially from the standard vacuum at those distances, which was derived, after all, based on a particular choice for the low energy action.  Thus relative to energy-momentum tensor of the standard vacuum, the additional structures can produce new short-distance divergences when we correspondingly evaluate the energy-momentum tensor in one of the effective states.  One property of these divergences, which will be shown in this section, is that they occur only at the initial-time hypersurface where the state is defined.  As a result, the counterterms chosen to cancel these divergences are purely geometric operators confined to this surface, 
\begin{equation}
S_{\rm surf} = \int_{\eta_0} d^3\vec x\, \sqrt{-h}\, {\cal L}_{\rm surf} , 
\label{gensurfS}
\end{equation}
where ${\cal L}_{\rm surf}$ is the counterterm Lagrangian.  

An important difference between this set of divergences is that they explicitly break some of the space-time symmetries that were still preserved by the standard vacuum.  This symmetry breaking can arise from many possible sources, depending upon what happens at the shortest of scales; but all of these possibilities are treated equivalently in the effective description of the state \cite{schalm,cliff}.  In the simplest case this symmetry breaking can arise because we have integrated out the dynamics of some excited field, coupled to the light field producing the inflation, to find the effective state.  Alternatively, the new structure in the state could also have its origin in a loss of classical, local Lorentz invariance at very short distances, where it could be replaced by a quantum-deformed symmetry or some other principle \cite{brandenberger,gary,transplanck,fate}.  Thus the counterterm Lagrangian can contain a more general set of geometrical objects, defined on the initial-time hypersurface, which are consistent with the symmetries left unbroken by the effective state we have chosen.

In the case of an isotropically expanding universe, with a spatially flat metric as in Eq.~(\ref{metric}), the set of surface tensors available includes those inherited from the bulk geometry---the metric, curvature and covariant derivative, $\{ g_{\mu\nu}, R^\sigma_{\ \lambda\mu\nu}, \nabla_\mu \}$---as well as invariants constructed using the normal $n^\mu$ and the induced metric along the surface, 
\begin{equation}
h_{\mu\nu} = g_{\mu\nu} - n_\mu n_\nu 
\label{induced}
\end{equation}
such as the extrinsic curvature, $K_{\mu\nu}$, defined by 
\begin{equation}
K_{\mu\nu} = h_\mu^{\ \lambda} \nabla_\lambda n_\nu . 
\label{extrinsic}
\end{equation}
Most generally, if the state also contains some spatial variation, we would include invariant counterterms constructed from the {\it induced\/} curvature $\hat R^\sigma_{\ \lambda\mu\nu}$, which is defined to be the Riemann tensor associated with the induced metric $h_{\mu\nu}$, as well as covariant surface derivatives.

In this section, we first illustrate the method for renormalizing the boundary divergences of a spatially flat effective state in Minkowski space.  In this case, the unique boundary counterterm is the surface tension; yet, by itself, this term is sufficient to absorb all of the divergences in the expectation value of the energy-momentum tensor.  It is also most straightforward in Minkowski space to show that the divergences do indeed occur precisely at the initial time hypersurface since all of the momentum integrals can be performed explicitly.  

We then extend to an isotropically expanding space-time.  The renormalization of the new divergences in the energy-momentum tensor in this background becomes more subtle---especially when the gravitational component of the theory is treated classically---because the counterterms must contain derivatives of the scale factor, which thus means that we need to include terms containing normal derivatives of some object at the surface.  

The trouble with such terms is that their effect on the classical gravitational contribution to the equations of motion can appear ill-defined.  For example, in varying the counterterm action with respect to a small change in the metric, $g_{\mu\nu} \to g_{\mu\nu} + \delta g_{\mu\nu}$, we encounter terms where a normal derivative acts on the variation, 
\begin{equation}
n^\lambda \nabla_\lambda \delta g_{\mu\nu} . 
\label{badnormal}
\end{equation}
When such a term is evaluated on the boundary, as are all of the state-dependent counterterms, the normal derivative cannot be removed by an integration by parts.  To provide a simple introduction to the renormalization of these boundary divergences produced by an effective initial state, in this article we shall follow a more standard approach, treating gravity always classically, noting the limitations of this approach when we encounter them.  

Ultimately, since the new features in the propagator represent the quantum interference between the initial state and a point source propagating at some later time, the correct approach for treating the new initial-time divergences is not to reduce the field theory side of Einstein's equation to a purely classical quantity, by taking the classical expectation value of the energy-momentum tensor, but is instead to treat the gravitational side of the equation as a quantum field theory evaluated at tree level.  We shall follow this approach later in \cite{next}.  In this treatment, the equations of motion arise as a renormalization condition that requires the vanishing of the tadpole graphs and the time-evolution of the matrix element provides an additional time integral which allows normal derivatives---the analogues in this setting of features like Eq.~(\ref{badnormal})---to be integrated by parts.  The cancellation of the new divergences then proceeds very similarly to that of \cite{dSgreens}, where normal derivatives acted on the scalar field.

\subsection{Minkowski space}

An elegant illustration of the renormalization of the boundary divergences resulting from an effective initial state occurs in Minkowski space, where there is a unique boundary counterterm, the surface tension.  Although only a single counterterm is available, it is sufficient to absorb all of the short-distance divergences produced by the energy and the momentum contained in a general effective state.  Part of the purpose in examining Minkowski space is to provide the simplest example of the renormalization of the new divergences associated with the effective state, but a second purpose is to see that the divergences are genuinely confined to the initial-time hypersurface.

Before analyzing these divergences, we first derive the contribution from a surface tension to the gravitational equations of motion.  The action for such a term, 
\begin{equation}
S_{\sigma} = \int d^3\vec x\, \sqrt{-h}\, \sigma , 
\label{tension}
\end{equation}
under a general variation of the metric changes by 
\begin{equation}
\delta S_{\sigma} = \int d^3\vec x\, \sqrt{-h}\, \bigl\{ 
{\textstyle{1\over 2}} \sigma h^{\mu\nu}\, \delta h_{\mu\nu} 
\bigr\} . 
\label{dtension}
\end{equation}
In these expressions, $h$ represents the determinant of the induced metric, defined in Eq.~(\ref{induced}) and the variation of this metric is a part of the variation of the full metric, 
\begin{equation}
\delta g_{\mu\nu} = \delta h_{\mu\nu} + n_\mu\, \delta n_\nu + n_\nu\, \delta n_\mu ,
\label{deltagdeltah}
\end{equation}
neglecting terms that are second order in the variation.  Since $n_\mu$ is orthogonal to the surface, $n^\mu h_{\mu\nu}=0$, we can rewrite 
\begin{equation}
h^{\mu\nu}\, \delta h_{\mu\nu} 
= h^{\mu\nu}\, \bigl( \delta g_{\mu\nu} - n_\mu\, \delta n_\nu - n_\nu\, \delta n_\mu \bigr) 
= h^{\mu\nu}\, \delta g_{\mu\nu} , 
\label{dhtodg}
\end{equation}
so that the variation of the surface action can be expressed as a variation of the full, four-dimensional metric, 
\begin{equation}
\delta S_{\sigma} = \int d^3\vec x\, \sqrt{-h}\, \delta g_{\mu\nu} 
\bigl\{ {\textstyle{1\over 2}} \sigma h^{\mu\nu} \bigr\} ,
\label{dtensiong}
\end{equation}
which contributes thus to the equation of motion, 
\begin{equation}
{2\over\sqrt{-h}} {\delta S_{\sigma}\over\delta g^{\mu\nu}} 
= \sigma\, h_{\mu\nu} . 
\label{dStensedg}
\end{equation}
The important property of this contribution is that its temporal components vanish, reducing to 
\begin{equation}
\cases{0 &for $\mu,\nu=0$\cr 
- \sigma\, \delta_{ij} &for $i,j\not= 0$ .\cr}
\label{flattense}
\end{equation}
The divergences in the energy-momentum tensor have exactly this form.

We next consider the effect of the scalar field.  To distinguish the standard vacuum modes of Minkowski space from their analogues in a curved background, we shall alter our notation slightly, denoting the Minkowski modes by $u_k(t)$ and writing the space-time coordinates more conventionally as $(t,\vec x)$ since there is no need to distinguish a conformal time coordinate.  The standard vacuum modes have the form
\begin{equation}
u_k(t) = {e^{-i\omega_k(t-t_0)}\over\sqrt{2\omega_k}} 
\label{flatmodes}
\end{equation}
where now, since the scale factor is trivial ($a(t)=1$), $\omega_k$ reduces to the standard frequency,
\begin{equation}
\omega_k = \sqrt{k^2 + m^2} . 
\label{minkfreq}
\end{equation}
Substituting these mode functions into the expressions for the non-bulk contributions to the energy density and the pressure provided in Eqs.~(\ref{rhofull})--(\ref{pfull}), and again setting $a=1$ in those expressions, yields
\begin{eqnarray}
\rho_{\rm surf}(t) 
&\!\!\!=\!\!\!& 0 
\nonumber \\
p_{\rm surf}(t) 
&\!\!\!=\!\!\!& - {1\over 6} \int {d^3\vec k\over(2\pi)^3}\, 
f_k^* \biggl[ 2 \omega_k + {m^2\over\omega_k} \biggr] 
e^{-2i\omega_k(t-t_0)} . 
\qquad
\label{flatsurf}
\end{eqnarray}
Already in this form we can see that the divergences have the same structure as a surface tension, Eq.~(\ref{flattense})---provided the oscillatory factor effectively confines the divergence to where its phase vanishes, at $t=t_0$.

Let us consider a more explicit form for the initial state by describing its structure through a power series in the frequency, 
\begin{equation}
f_k = \sum_{n=0}^\infty c_n {m^n\over\omega_k^n} 
+ \sum_{n=1}^\infty d_n {\omega_k^n\over M^n} . 
\label{minkfk}
\end{equation}
Here we have used the frequency as our expansion variable, rather than the spatial momentum $k$, since the former is finite at $k=0$, so that the terms in the first series do not produce any unnecessary divergences while at short distances the difference between $\omega_k$ and $k$ becomes increasingly negligible.  The first of these sums is less interesting as a parameterization of the trans-Planckian problem since all of these terms vanish in the $k\to\infty$ limit, except for the marginal, $n=0$, term.  The second sum is negligible at long distances if assume that $m\ll M$---that it does not vanish altogether is an artifact of having chosen the frequency as the expansion variable---but grows in significance above $k\sim M$.  We shall analyze both of these sets of terms, although our principal interest in is the second, trans-Planckian component.

The general structure of the contributions from each of these terms to the pressure is similar, being ultimately a product of poles and Hankel functions, which can be arranged, through standard relations among these functions of different orders, to contain only the zeroth and first Hankel functions accompanied by poles whose order is transparently related to the degree of divergence of the momentum integral.  As an example, consider the lowest order divergence, coming from the $n=4$ term in the first series.  Inserting it into the expression for the pressure yields 
\begin{equation}
p_{\rm surf}(t) 
= - {c_4^* m^4\over 6} \int {d^3\vec k\over (2\pi)^3}\, 
\biggl[ {2\over\omega_k^3} + {m^2\over\omega_k^5} \biggr] e^{-2i\omega_k(t-t_0)} ; 
\label{c4div}
\end{equation}
the second term in the integrand is manifestly finite while the first one diverges logarithmically at $t=t_0$.  Integrating over the momenta we find 
\begin{eqnarray}
p_{\rm surf}(t) 
&\!\!\!=\!\!\!& {ic_4^* m^4\over 12\pi} H_0^{(2)}[2m(t-t_0)] + {\rm finite}
\nonumber \\
&\!\!\!=\!\!\!& {c_4^* m^4\over 6\pi^2} \ln\bigl[ m(t-t_0) \bigr] 
+ {\rm finite}
\label{c4divH2}
\end{eqnarray}
which explicitly shows that this state-dependent contribution to the pressure is finite everywhere except at $t=t_0$ where it has the logarithmic divergence that we anticipated.

This pattern persists when we consider higher order terms in the initial state structure function.  For example, proceeding to include the $n=2$ and $n=0$ terms,
\begin{eqnarray}
p_{\rm surf}(t) 
&\!\!\!=\!\!\!& - {1\over 6} \int {d^3\vec k\over (2\pi)^3}\, 
\biggl[ c_0^* + c_2^*{m^2\over\omega_k^2} + c_4^*{m^4\over\omega_k^4} \biggr] 
\nonumber \\
&&\qquad\qquad\times
\biggl[ 2\omega_k + {m^2\over\omega_k} \biggr] e^{-2i\omega_k(t-t_0)} 
\quad
\label{testdiv}
\end{eqnarray}
we find, 
\begin{eqnarray}
&&\!\!\!\!\!\!\!\!\!\!\!\!
p_{\rm surf}(t) 
\nonumber \\
&\!\!\!=\!\!\!& {im^4\over 24\pi} \biggl\{ 
c_0^* \biggl[ 
{3\, H_1^{(2)}[2m(t-t_0)]\over 2m^3(t-t_0)^3}
- {5\, H_1^{(2)}[2m(t-t_0)]\over 2m(t-t_0)}
\nonumber \\
&&\qquad\quad
- {3\, H_0^{(2)}[2m(t-t_0)]\over 2m^4(t-t_0)^2}
+ 2\, H_0^{(2)}[2m(t-t_0)] \biggr] 
\nonumber \\
&&\quad
+\ c_2^* \biggl[ 
{H_1^{(2)}[2m(t-t_0)]\over m(t-t_0)}
- 3\, H_0^{(2)}[2m(t-t_0)] \biggr] 
\nonumber \\
&&\quad
+\ 2 c_4^*\, H_0^{(2)}[2m(t-t_0)] 
\biggr\} . 
\label{testdivH2}
\end{eqnarray}
Noting that the Hankel functions only diverge when their arguments vanish, 
\begin{eqnarray}
H_0^{(2)}[2m(t-t_0)] 
&\!\!\!=\!\!\!& 
- {2i\over\pi}\ln\bigl( m(t-t_0) \bigr) - {2i\over\pi} \gamma + 1
\nonumber \\
&&
+\ {\cal O} \bigl( (t-t_0)^2 \bigr) 
\nonumber \\
H_1^{(2)}[2m(t-t_0)] 
&\!\!\!=\!\!\!& 
{i\over\pi} {1\over m(t-t_0)} + {\cal O} \bigl( (t-t_0) \bigr) , 
\label{Hankels}
\end{eqnarray}
we discover that the divergences occur where we expect, at the initial time, with the order we expect, quadratically for the $c_2$ term and quartically for the $c_0$ marginal term.  The behavior of a generic trans-Planckian term is completely analogous, with a leading $(t-t_0)^{-n-4}$ pole being associated with the $d_n$ moment.

Having shown that the new structure only leads to initial-time divergences, it is convenient to proceed next by first evaluating the terms at $t=t_0$ and then dimensionally regularizing the spatial momentum integrals, by extending the number of dimensions to $3-2\epsilon$, to extract the poles.  This regularization method then treats all divergences as $1/\epsilon$ poles and moreover yields a finite result for all of the odd-$n$ terms in either sum in Eq.~(\ref{minkfk}).

Following this approach for the full structure function in Eq.~(\ref{minkfk}), and applying the dimensional regularization described in Eq.~(\ref{dreps}), the divergences in the pressure at $t=t_0$ are 
\begin{eqnarray}
p_{\rm surf}(t_0) 
&\!\!\!=\!\!\!& {m^4\over 32\pi^2} \biggl[ c_0^* - {8\over 3} c_4^* \biggr] 
\biggl[ {1\over\epsilon} - \gamma + \ln{4\pi\mu^2\over m^2} \biggr] 
\nonumber \\
&&
+\ {m^4\over 16\pi^2} \sum_{n=1}^\infty d_{2n}^* 
\biggl( {m^2\over M^2} \biggr)^n 
{(n+1)(2n-1)!!\over 2^n (n+2)!} 
\nonumber \\
&&\qquad\qquad
\times\biggl[ {1\over\epsilon} + \psi(n+3) + \ln{4\pi\mu^2\over m^2} \biggr] 
\nonumber \\
&& 
+\ {\rm finite} , 
\label{psurfM}
\end{eqnarray} 
where $\psi(n+3)$ is the digamma function.  This divergence can be cancelled by including a surface tension of the form 
\begin{eqnarray}
\sigma 
&\!=\!& - {m^4\over 16\pi^2} {1\over\epsilon} \biggl\{ 
- {4\over 3} c_4^* + {1\over 2} c_0^* 
\nonumber \\
&&
+\ \sum_{n=1}^\infty d_{2n}^* 
\biggl( {m^2\over M^2} \biggr)^n 
{(n+1)(2n-1)!!\over 2^n (n+2)!} 
\biggr\}
+ {\rm finite} ,
\nonumber \\
&& 
\label{sigmaM}
\end{eqnarray} 
where the finite terms depend on the regularization scheme.  Setting them to zero corresponds to a minimal subtraction (MS) scheme but we could use them to cancel some of the finite artifacts left from the dimensional regularization, which generalizes the familiar $\overline{\rm MS}$ scheme usually applied.

\subsection{An isotropically expanding background}

The basic philosophy for the boundary renormalization in a curved space-time is the same as that of Minkowski space, although it is complicated by the much richer set of non-vanishing boundary invariants available.  We already saw an analogous growth in the complexity while renormalizing the bulk properties of gravity; in Minkowski space there is only the renormalization of the cosmological constant while in an expanding background we also have the renormalization of the coefficients of the two and four derivative terms.  Similarly, while a surface tension was sufficient to absorb the boundary divergences of Minkowski space, higher derivative terms in the boundary action will be required to cancel the divergences in an expanding background.  In principle, when considering the most general set of trans-Planckian features of a state, we would require an infinite set of counterterms; but their effects will be suppressed by powers\footnote{Here we have broadly characterized the suppression by powers of $H/M$ although other scales, given by derivatives of $H$, are available.  In the slowly rolling limit of inflation, $H'\ll H^2$, so these derivative scales are usually already suppressed and at any given order the $(H/M)^n$ terms have the largest effect.} of $H/M$, so that for a given measurement only the leading trans-Planckian effects will be significant.

There is a second difficulty, already mentioned in the beginning of this section, that results from the need to include normal derivatives, which here correspond to $\eta$-derivatives, in the boundary action.  To treat these terms with a purely classical description of gravity, we need to restrict the class of allowed variations of the metric at the boundary.  For example, in deriving the coefficient of the Gibbons-Hawking term \cite{gh}, it is ordinarily assumed \cite{wald} that $\delta g_{\mu\nu}(\eta_0) = 0$ but that $n^\lambda \nabla_\lambda \delta g_{\mu\nu}(\eta_0) \not= 0$, whereas here we shall impose exactly the opposite constraint.  When we do so, we obtain precisely the correct structure to cancel the next divergence beyond that proportional to the surface tension.

The origin of this peculiarity lies in an incompatibility between the broken symmetries of the quantum theory and the classical symmetries of gravity.  A completely general effective state breaks Lorentz invariance at short distances whereas the classical description of gravity always implicitly assumes that over infinitesimal distances the space-time looks locally flat.  Thus any difficulties in absorbing the energy-momentum divergences by renormalizing the gravitation action only occur because we have not allowed for a sufficiently general set of operators in the geometric component of the theory to account for these symmetries broken by the quantum state of the scalar field.

The proper treatment therefore is to evaluate the gravitational part of the theory quantum mechanically as well so that both the scalar field and geometric components of the theory are on the same footing \cite{next}.  However, it is still instructive to follow a derivation that more closely approximates the standard classical derivation, noting the constraints we must impose on the variation of the metric at the boundary to obtain the correct contributions from the boundary counterterms to cancel the new divergences from the scalar field.  Since the available counterterms at each order in a derivative expansions proliferate very rapidly, we shall illustrate the approach in detail only for the first few classes of divergences before commenting more generally on the structure of the counterterms required for absorbing the divergences from the leading trans-Planckian signal.  Finally, we shall show that after being renormalized, the contribution to the energy and the momentum from the trans-Planckian aspects of a state are very small in comparison with the vacuum energy that drives inflation.

\subsubsection{Boundary divergences in the energy-momentum tensor}

Recall that the state-dependent contributions to the total energy density and the pressure, given in Eqs.~(\ref{rhofull})--(\ref{pfull}), are 
\begin{eqnarray}
\rho_{\rm surf} 
&\!\!\!=\!\!\!& {1\over 2} {1\over a^2} \int {d^3\vec k\over(2\pi)^3}\, 
f_k^* \bigl[ U'_k U'_k + (k^2+a^2m^2) U_k U_k \bigr]
\nonumber \\
p_{\rm surf} 
&\!=\!& \rho_{\rm surf} 
- {1\over a^2} \int {d^3\vec k\over(2\pi)^3}\, 
f_k^* \biggl[ {2\over 3} k^2 + a^2m^2 \biggr] U_k U_k . 
\nonumber \\
&&
\label{rhopsurf}
\end{eqnarray}
Substituting the form of the vacuum modes as in Eq.~(\ref{adiabat}) into these equations, they become of a form to which we can more readily apply our adiabatic expansion, 
\begin{eqnarray}
\rho_{\rm surf} 
&\!\!\!=\!\!\!& {1\over 4} {1\over a^4} \int {d^3\vec k\over(2\pi)^3}\, 
{e^{-2i\int_{\eta_0}^\eta d\eta'\, \Omega_k(\eta')}\over\Omega_k} 
f_k^* 
\nonumber \\
&&\qquad\times
\biggl[ 
{a^{\prime\prime}\over a} + \left( {a'\over a} \right)^2 
+ {a'\over a} {\Omega'_k\over\Omega_k} 
+ {1\over 2} {\Omega_k^{\prime\prime}\over\Omega_k} 
- {1\over 2} \left( {\Omega_k^\prime\over\Omega_k} \right)^2 
\nonumber \\
&&\qquad\quad
+\ 2 i \Omega_k {a'\over a} + i \Omega'_k
\biggr]
\label{rhosurfO}
\end{eqnarray}
and
\begin{eqnarray}
p_{\rm surf} 
&\!\!\!=\!\!\!& {1\over 4} {1\over a^4} \int {d^3\vec k\over(2\pi)^3}\, 
{e^{-2i\int_{\eta_0}^\eta d\eta'\, \Omega_k(\eta')}\over\Omega_k} 
f_k^* 
\nonumber \\
&&\qquad\times
\biggl[ 
-\ {4\over 3} \Omega_k^2 
- {1\over 3} {a^{\prime\prime}\over a} 
+ \left( {a'\over a} \right)^2 
- {2\over 3} a^2m^2 
\nonumber \\
&&\qquad\quad
+\ {a'\over a} {\Omega'_k\over\Omega_k} 
- {1\over 6} {\Omega_k^{\prime\prime}\over\Omega_k} 
+ {1\over 2} \left( {\Omega_k^\prime\over\Omega_k} \right)^2 
\nonumber \\
&&\qquad\quad
+\ 2 i \Omega_k {a'\over a} + i \Omega'_k
\biggr] . 
\label{psurfO}
\end{eqnarray}
Once we consider a particular analytical form for the initial state structure, $f_k$, we can then extract the scale dependence of the divergences by evaluating these functions at the initial time, $\eta=\eta_0$, and dimensionally regularizing the remaining momentum integrals.  The form we shall use for this structure is 
\begin{equation}
f_k = \sum_{n=0}^\infty c_n \biggl( {am\over\omega_k} \biggr)^n 
+ \sum_{n=1}^\infty d_n \biggl( {\omega_k\over aM} \biggr)^n , 
\label{fkfrw}
\end{equation}
where $\omega_k$ is again 
\begin{equation}
\omega_k = \sqrt{k^2+a^2m^2} . 
\label{omegaenc}
\end{equation}
Note that while we have included the scale factor in the expressions for $f_k$ and $\omega_k$ they are not strictly necessary since they do not have actually any time dependence, being always evaluated at $\eta=\eta_0$.  Note further that we have again chosen a particular function for our expansion parameter in these series since this choice considerably simplifies the momentum integrals we shall encounter; in principle we could have chosen any other function with the same behavior in the $k\to\infty$ limit which we are studying.  One artifact of this choice is that, as in the Minkowski example before, the trans-Planckian features will have a little lingering long-distance structure, which we can cancel by an appropriate choice of other terms in these series.

\subsubsection{Two relevant examples}

As an illustration, let us examine the two simplest cases in detail.  Counting powers of $k$ in the ultraviolet limit of the momentum integral in Eq.~(\ref{psurfO}), these cases correspond to the divergences arising from the relevant boundary conditions, 
\begin{equation}
f_k = c_4 {a^4m^4\over\omega_k^4}\quad{\rm and}\quad 
c_3 {a^3m^3\over\omega_k^3} ;
\label{twocases}
\end{equation}
that these should both be relevant, or long-distance, modifications of the state should not be surprising since even in the bulk renormalization, the operators for a free scalar field, which are all relevant or marginal, produced divergences in the energy-momentum tensor.  We shall denote the contributions to the energy density from each of these structures by $\rho_{\rm surf}^{(4)}$ and $\rho_{\rm surf}^{(3)}$, with a similar notation for the pressure.  

Evaluating the density (\ref{rhosurfO}) and the pressure (\ref{psurfO}) at the initial time and using the adiabatic expansion of their integrands given in detail in Eqs.~(\ref{rhosurfA4})--(\ref{psurfA4}) of the Appendix, we obtain the following expressions for the divergences for each of these two cases, 
\begin{eqnarray}
\rho_{\rm surf}^{(4)}(\eta_0)
&\!\!\!=\!\!\!& {\rm finite} 
\nonumber \\ 
p_{\rm surf}^{(4)}(\eta_0)
&\!\!\!=\!\!\!& - {c_4^*m^4\over 12\pi^2} 
\biggl[ {1\over\epsilon} + \ln{\mu^2\over m^2} \biggr]
+ \cdots
\label{rhopcase1}
\end{eqnarray}
and
\begin{eqnarray}
\rho_{\rm surf}^{(3)}(\eta_0)
&\!\!\!=\!\!\!& {ic_3^*m^3\over 8\pi^2} {a'\over a^2} 
\biggl[ {1\over\epsilon} + \ln{\mu^2\over m^2} \biggr]
+ \cdots 
\nonumber \\ 
p_{\rm surf}^{(3)}(\eta_0)
&\!\!\!=\!\!\!& {ic_3^* m^3\over 8\pi^2} {a'\over a^2} 
\biggl[ {1\over\epsilon} + \ln{\mu^2\over m^2} \biggr]
+ \cdots . 
\label{rhopcase2}
\end{eqnarray}
Here we have not written any of the finite contributions---other than those depending on the renormalization scale $\mu$---such as factors of $-\gamma$ or $\ln 4\pi$, which are the usual artifacts of dimensional regularization.

The first of these examples has the same structure as the variation of a surface tension term.  If we start again from the action, 
\begin{equation}
S_\sigma = \int d^3\vec x\, \sqrt{-h}\, \sigma , 
\label{tension2}
\end{equation}
its variation, 
\begin{equation}
{2\over\sqrt{-h}} {\delta S_\sigma\over\delta g^{\mu\nu}} 
= \sigma\, h_{\mu\nu} , 
\label{dStension}
\end{equation}
contributes the following to the equations of motion at the boundary, 
\begin{equation}
\sigma\, h_\mu^{\ \nu} = \cases{0 &for $\mu,\nu=0$\cr \sigma\, \delta_i^{\ j} &for $\mu,\nu\not= 0$\cr} ,
\label{dStensiondu}
\end{equation}
precisely matching the space-time structure of the divergences in Eq.~(\ref{rhopcase1}) that only occur in the pressure.

The next example is more interesting since it provides the first and simplest example that contains normal derivatives evaluated at the boundary.  The unique one-derivative boundary invariant available is the trace of the extrinsic curvature, $K=h^{\mu\nu}K_{\mu\nu}$, 
\begin{equation}
S_\kappa = \kappa \int d^3\vec x\, \sqrt{-h}\, K . 
\label{Sextr}
\end{equation}
The variation of this action with respect to a small change in the full metric, $\delta g_{\mu\nu}$, is 
\begin{eqnarray}
\delta S_\kappa 
&\!\!\!=\!\!\!& 
\kappa\, \int d^3\vec x\, \sqrt{-h}\, \Bigl\{ 
\delta g_{\mu\nu} {\textstyle{1\over 2}} 
\bigl[ h^{\mu\nu} - n^\mu n^\nu \bigr]\, K 
\nonumber \\
&&\qquad
+\ h^{\mu\nu} \nabla_\mu \bigl( h_{\nu\lambda}\, \delta n^\lambda \bigr) 
+ {\textstyle{1\over 2}} h^{\mu\nu} n^\lambda \nabla_\lambda \delta g_{\mu\nu} 
\Bigr\} ; 
\nonumber \\
&&
\label{dSextr}
\end{eqnarray}
in deriving this variation, we have simplified the structure of the integrand by applying 
\begin{eqnarray}
n^\mu n^\nu\, \delta h_{\mu\nu} &\!\!\!=\!\!\!& 0
\nonumber \\
n^\lambda\, \delta h_{\lambda\nu} &\!\!\!=\!\!\!& 
- h_{\nu\lambda}\, \delta n^\lambda
\label{surfrels}
\end{eqnarray}
which follow respectively from the preservation of the unit length of the normal and its orthogonality to the induced metric under a general variation, 
\begin{eqnarray}
(n_\mu + \delta n_\mu) (n_\nu + \delta n_\nu) 
(g_{\mu\nu} + \delta g_{\mu\nu}) &\!\!\!=\!\!\!& 1
\nonumber \\
(n^\lambda + \delta n^\lambda) (h_{\lambda\nu} + \delta h_{\lambda\nu}) &\!\!\!=\!\!\!& 0 . 
\label{surfrelsderv}
\end{eqnarray}

The second term in the expression for $\delta S_\kappa$ is a total derivative with respect to the coordinates {\it along\/} the surface \cite{he}; the appearances of the induced metric act to project the tensors in this term into the space of surface tensors.  Thus this piece does not have any role in the surface equations of motion.  The third term represents a normal derivative of a variation of the metric.  In the derivation of the Gibbons-Hawking term, we would use just this term to cancel a similar term coming from the integration by parts of a term from the variation of the curvature, $\delta R$.  Here we must instead constrain our variations of the bulk metric to those that satisfy the condition, 
\begin{equation}
n^\lambda \nabla_\lambda \delta g_{\mu\nu} \bigr|_{\eta=\eta_0} = 0 , 
\label{Nconstraint}
\end{equation}
since otherwise we would need also to renormalize the coefficient of the bulk curvature term, whereas we already know that the state-dependent divergences only occur at the boundary.  As mentioned before, the true resolution to this obstruction is to treat the gravitational component of the theory quantum mechanically as well \cite{next}; but for now, to obtain a broad idea of how the renormalization of the boundary divergences proceeds, we restrict to this class of variations.  

The remaining contribution to the equations of motion at the boundary from the extrinsic curvature action is 
\begin{equation}
{2\over\sqrt{-h}} {\delta S_\sigma\over\delta g^{\mu\nu}} 
= \kappa\, K\, \bigl( h_{\mu\nu} - n_\mu n_\nu \bigr) . 
\label{dSkappa}
\end{equation}
In an expanding background, $K$ is given by 
\begin{equation}
K = 3H = {3a'\over a^2} . 
\label{KvalH}
\end{equation}
Substituting this form into the gravitational equations at the boundary, Eq.~(\ref{dSkappa}), we obtain 
\begin{equation}
\kappa\, K\, \bigl( h_\mu^{\ \nu} - n_\mu n^\nu \bigr)
= \cases{\ - 3\kappa {\displaystyle{a'\over a^2}} &for $\mu=\nu=0$\cr 
\quad 3\kappa {\displaystyle{a'\over a^2}}\, \delta_i^{\ j} &for $\mu,\nu\not= 0$\cr} ,
\label{dSkappadu}
\end{equation}
which is again precisely the form needed to cancel the second set of divergences in the energy density and pressure in Eq.~(\ref{rhopcase2}).  Note that the opposite signs in this gravitational side of the equation are needed since the pressure is itself defined with a minus sign, as in Eq.~(\ref{Texptdef}).

The procedure for determining the correct geometric counterterms at higher orders is similar to these examples, although it becomes more complicated by the greater number of terms available at a particular order in derivatives.  For example, at the next order, 
\begin{equation}
f_k = c_2 {a^2m^2\over\omega_k^2} , 
\label{lastcase}
\end{equation}
the boundary divergences contain two derivatives, 
\begin{eqnarray}
\rho_{\rm surf}^{(2)}(\eta_0)
&\!\!\!=\!\!\!& {c_2^*m^2\over 16\pi^2} {1\over a^2} 
\biggl[ {a^{\prime\prime}\over a^2} + \biggl( {a'\over a^2} \biggr)^2 \biggr]
\biggl[ {1\over\epsilon} + \ln{\mu^2\over m^2} \biggr]
+ \cdots 
\nonumber \\ 
p_{\rm surf}^{(2)}(\eta_0)
&\!\!\!=\!\!\!& {c_2^*m^2\over 16\pi^2} {1\over a^2} \biggl[ 
{1\over 3} {a^{\prime\prime}\over a^2} 
+ \biggl( {a'\over a^2} \biggr)^2 \biggr]
\biggl[ {1\over\epsilon} + \ln{\mu^2\over m^2} \biggr]
+ \cdots , 
\nonumber \\
&&
\label{rhopcase3}
\end{eqnarray}
and so should be absorbed by corresponding terms on the boundary.  Even for a spatially flat metric, the number of two derivative terms is considerably larger,\footnote{A sixth possibility, $n_\mu \nabla_\nu K^{\mu\nu}$, is equivalent to $-K_{\mu\nu}K^{\mu\nu}$.} 
\begin{equation}
K^2,\ K_{\mu\nu}K^{\mu\nu},\ n^\lambda \nabla_\lambda K,\ 
R,\ n^\mu n^\nu R_{\mu\nu}, 
\label{twoderivs}
\end{equation}
than at lower orders, where we had unique counterterms.

\subsubsection{Power counting, trans-Planckian signals and their backreaction}

From the preceding examples, we can describe the procedure for renormalizing the boundary divergences produced by an arbitrary trans-Planckian effective state.  Let us consider one of the terms in the trans-Planckian part of the structure function, 
\begin{equation}
f_k = d_n {\omega_k^n\over (aM)^n} . 
\label{fkeg}
\end{equation}
Evaluating the energy density and the pressure contributed by this part of the short-distance structure at the initial-time surface, we have from Eqs.~(\ref{rhosurfO})--(\ref{psurfO}) that 
\begin{eqnarray}
\rho_{\rm surf}(\eta_0) 
&\!\!\!=\!\!\!& {1\over 4} {1\over a^{n+4}} {d_n^* \over M^n} 
\int {d^3\vec k\over(2\pi)^3}\, 
\omega_k^n
\nonumber \\
&&
\times \biggl\{ 
{1\over\Omega_k} \biggl[ {a^{\prime\prime}\over a} + \left( {a'\over a} \right)^2 \biggr] 
\nonumber \\
&&\quad
+ {a'\over a} {\Omega'_k\over\Omega_k^2} 
+ {1\over 2} {\Omega_k^{\prime\prime}\over\Omega_k^2} 
- {1\over 2} {1\over\Omega_k} \left( {\Omega_k^\prime\over\Omega_k} \right)^2 
\nonumber \\
&&\quad
+\ 2 i {a'\over a} + i {\Omega'_k\over\Omega_k} 
\biggr\} 
\label{rhosurffk}
\end{eqnarray}
and
\begin{eqnarray}
p_{\rm surf}(\eta_0) 
&\!\!\!=\!\!\!& {1\over 4} {1\over a^{n+4}} {d_n^*\over M^n}
\int {d^3\vec k\over(2\pi)^3}\, 
\omega_k^n 
\nonumber \\
&&\times
\biggl\{  
-\ {4\over 3} \Omega_k 
+ {1\over\Omega_k} \biggl[ 
- {1\over 3} {a^{\prime\prime}\over a} 
+ \left( {a'\over a} \right)^2 
- {2\over 3} a^2m^2 
\biggr]
\nonumber \\
&&\quad
+\ {a'\over a} {\Omega'_k\over\Omega_k^2} 
- {1\over 6} {\Omega_k^{\prime\prime}\over\Omega_k^2} 
+ {1\over 2} {1\over\Omega_k} \left( {\Omega_k^\prime\over\Omega_k} \right)^2 
\nonumber \\
&&\quad
+\ 2 i {a'\over a} + i {\Omega'_k\over\Omega_k} 
\biggr\} . 
\label{psurffk}
\end{eqnarray}
Together, the $\omega_k^n\, d^3\vec k$ factor in both the integrands scales as $k^{n+3}$ for large values of the momentum and therefore to determine the divergences we must expand the remaining terms in the integrands to order $\omega^{-n-3}$ using our adiabatic approximation.  Doing so, the divergent parts of energy density and the pressure have the following general structures, 
\begin{eqnarray}
\rho_{\rm surf}(\eta_0) 
&\!\!\!=\!\!\!& {1\over 4} {1\over a^{n+4}} {d_n^* \over M^n} 
\int {d^3\vec k\over(2\pi)^3}\, 
\nonumber \\
&&
\times \biggl\{ 
A^{(\rho,n+4)}(a)\, {1\over\omega_k^3} + \cdots + A^{(\rho,1)}(a)\, \omega_k^n 
\biggr\} 
\nonumber \\
&&
+\ {\rm finite}
\nonumber \\
p_{\rm surf}(\eta_0) 
&\!\!\!=\!\!\!& {1\over 4} {1\over a^{n+4}} {d_n^*\over M^n}
\int {d^3\vec k\over(2\pi)^3}\, 
\nonumber \\
&&\times
\biggl\{  
A^{(p,n+4)}(a)\, {1\over\omega_k^3} + \cdots + A^{(p,0)}(a)\, \omega_k^{n+1} 
\biggr\} 
\nonumber \\
&&
+\ {\rm finite} . 
\label{rhopsurfabst}
\end{eqnarray}
In these equations we have used $A^{(\rho,i)}(a)$ (or $A^{(p,i)}(a)$) to represent the scale-dependent prefactor we obtain, which can contain up to $i$ derivatives\footnote{As seen in the appendix, in Eqs~(\ref{rhosurfA4})--(\ref{psurfA4}), some of the components of this prefactor might contain powers of $a^2m^2$ at the expense of the derivatives so that the prefactor more generally contains terms of the structure, $\bigl\{ (i-2j)\ {\rm derivatives\ of}\ a \bigr\} \times \bigl( a^2 m^2 \bigr)^j$.} of the scale factor $a(\eta)$.  

With dimensional regularization, only those integrands that depend on an odd power of $\omega_k$ actually diverge, since we are here integrating only over the three spatial dimensions.  Therefore, upon regularizing the momentum integrals and applying the general formula of Eq.~(\ref{dreps}), we obtain the following poles, 
\begin{eqnarray}
&&\!\!\!\!\!\!\!\!\!\!\!
\rho_{\rm surf}(\eta_0) 
\nonumber \\
&\!\!\!=\!\!\!& {d_n^*\over 16\pi^2} {1\over\epsilon} \biggl[ 
{\hat A^{(\rho,n+4)}(a)\over a^{n+4}M^n} 
- {1\over 2} {m^2\over M^2} {\hat A^{(\rho,n+2)}(a)\over a^{n+2}M^{n-2}} 
+ \cdots 
\biggr] 
\nonumber \\
&&\!\!\!\!\!\!\!\!\!\!\!
p_{\rm surf}(\eta_0) 
\nonumber \\
&\!\!\!=\!\!\!& {d_n^*\over 16\pi^2} {1\over\epsilon} \biggl[  
{\hat A^{(p,n+4)}(a)\over a^{n+4}M^n} 
- {1\over 2} {m^2\over M^2} {\hat A^{(p,n+2)}(a)\over a^{n+2}M^{n-2}} 
+ \cdots 
\biggr] ; 
\nonumber \\
&&
\label{rhopsurfabsDR}
\end{eqnarray}
the prefactors, $\hat A^{(\rho,i)}(a)$ and $\hat A^{(p,i)}(a)$, are now functions of the scale factor only and contain exactly $i$ derivatives of $a(\eta)$, since the parts that depended on the mass $m$ properly belong to higher order effects, once we have integrated over the momenta.  Very generally then, these factors scale at worst as 
\begin{equation}
\hat A^{(\rho,i)}(a), \hat A^{(p,i)}(a) \propto (aH)^i
\label{hatAslow}
\end{equation}
in the slowly rolling limit of inflation where derivatives of $H$ are small.  Applying this scaling behavior to the generic expressions for the regularized density and pressure, we obtain the following leading contribution in the $H \gg m$ limit, 
\begin{eqnarray}
\rho_{\rm surf}(\eta_0) 
&\!\!\!\propto\!\!\!& {H^4\over 16\pi^2} {H^n\over M^n} {d_n^*\over\epsilon} \Bigl[ 1 + {\cal O}\bigl( m^2/H^2 \bigr) + {\cal O}\bigl( H'/H^2 \bigr) \Bigr] 
\nonumber \\
p_{\rm surf}(\eta_0) 
&\!\!\!\propto\!\!\!& {H^4\over 16\pi^2} {H^n\over M^n} {d_n^*\over\epsilon} 
\Bigl[ 1 + {\cal O}\bigl( m^2/H^2 \bigr) + {\cal O}\bigl( H'/H^2 \bigr) \Bigr] ; 
\nonumber \\
&&
\label{rhopsurfabsDRH}
\end{eqnarray}
note that the constants of proportionality are not necessarily equal for these two relations.  Applying a renormalization scheme such as $\overline{\rm MS}$, the remaining finite contributions of the trans-Planckian parts of the effective state are thus generically of the order, 
\begin{equation}
\rho^R_{\rm surf}, p^R_{\rm surf}
\sim {H^4\over 16\pi^2} {H^n\over M^n} d_n^* , 
\label{rhopsurfabsMS}
\end{equation}
for the $n^{\rm th}$ order moment of the structure function.

Let us compare these renormalized contributions to the energy-momentum tensor with the usual vacuum energy density necessary to drive an inflationary expansion.  To leading order, the vacuum energy can be approximated by the de Sitter limit where $H$ is nearly constant, 
\begin{equation}
\rho_{\rm vac} \sim - p_{\rm vac} \sim M_{\rm pl}^2 H^2 . 
\label{inflvac}
\end{equation}
Comparing this vacuum energy to the typical scale for the $n^{\rm th}$ moment of a trans-Planckian effective state, we see that the backreaction is quite small, 
\begin{equation}
{\rho^R_{\rm surf}\over \rho_{\rm vac}}
\sim {1\over 16\pi^2} {H^2\over M_{\rm pl}^2} {H^n\over M^n} , 
\label{backreact}
\end{equation}
up to factors of order one.

\subsubsection{The leading trans-Planckian signal}

We conclude this section by showing in a little more detail the divergences, and consequently the scale-dependence of the finite part as well, of the leading trans-Planckian part of the effective state, 
\begin{equation}
f_k = d_1 {\omega_k\over aM} . 
\label{firstmoment}
\end{equation}
In this case, the explicit structure of the boundary divergences is 
\begin{eqnarray}
&&\!\!\!\!\!\!\!\!
\rho_{\rm surf}(\eta) = p_{\rm surf}(\eta_0)
\nonumber \\
&\!\!\!=\!\!\!& 
{1\over 128\pi^2} {id_1^*\over M} {1\over\epsilon}
\biggl\{ 
- 6 m^4 {a'\over a^2} 
+ {a^{\prime\prime\prime\prime\prime}\over a^6}  
- 3 {a^{\prime\prime\prime\prime}\over a^5} {a'\over a^2} 
- 8 {a^{\prime\prime\prime}\over a^4} {a^{\prime\prime}\over a^3} 
\nonumber \\
&&\qquad
+ 6 {a^{\prime\prime\prime}\over a^4} \left( {a'\over a^2} \right)^2 
+ 10 \left( {a^{\prime\prime}\over a^3} \right)^2 {a'\over a^2} 
- 6 {a^{\prime\prime}\over a^3} \left( {a'\over a^2} \right)^3  
\biggr\}
\nonumber \\
&& 
+\ {\rm finite} . 
\label{rhopsurftP1}
\end{eqnarray}
In terms of the Hubble scale $H=a'/a^2$, this result can also be re-expressed as 
\begin{eqnarray}
&&\!\!\!\!\!\!\!
\rho_{\rm surf}(\eta) = p_{\rm surf}(\eta_0)
\nonumber \\
&\!\!\!=\!\!\!& 
{1\over 128\pi^2} {id_1^*\over M} {1\over\epsilon}
\biggl\{ 
- 6 m^4 H 
+ 16 H^5
+ 58 {H'\over a} H^3 
+ 26 {H^{\prime\prime}\over a^2} H^2 
\nonumber \\
&&\qquad
+ 34 {(H')^2\over a^2} H
+ 12 {H^{\prime\prime}\over a^2} {H'\over a} 
+ 7 {H^{\prime\prime\prime}\over a^3} H 
+ {H^{\prime\prime\prime\prime}\over a^4}
\biggr\}
\nonumber \\
&& 
+\ {\rm finite} . 
\label{rhopsurftP1H}
\end{eqnarray}
Upon renormalizing this divergence, and again considering the $H^2\gg H',m^2$ limit which corresponds to the slowly rolling regime of inflation, we see that the backreaction from source of the leading trans-Planckian signal in the effective state scales only as 
\begin{equation}
\rho_{\rm surf} = p_{\rm surf} \sim 
{H^4\over 8\pi^2} {H\over M}\, (id_1^*) + \cdots . 
\label{leadtPH}
\end{equation}
Note that the appearance of the factor of $i$ means that $d_1$ should itself be purely imaginary so that the boundary counterterm action remains purely real.  This behavior is consistent with what we found when examining the loop corrections for an interacting scalar field theory evaluated in an effective initial state \cite{greens,dSgreens}.

\section{Conclusion}
\label{conclude}

The basic idea of the effective theory of an initial state is that a discrepancy can exist between what is the true state of the system and the state we have chosen to use in a quantum field theory, which thereby defines the propagator and the matrix elements of operators.  Over distances where we have a good empirical understanding of nature and a reasonable knowledge about the relevant dynamics, we can usually make an appropriate choice for this state.  Yet there always exist shorter distances where the behavior of nature is unknown and the correct state might not match with that we obtained by extrapolating our understanding at long distances down to these much shorter scales.  This discrepancy is particularly important for inflation, where the relevant fields and their dynamics have not been observed directly and where the natural energy scale, the Hubble scale $H$, can be an appreciable fraction of the Planck scale, $M_{\rm pl}$.  At this scale, gravity becomes strongly interacting and so we do not even have a predictive understanding of the behavior of space-time.  Since inflation naturally produces a set of primordial perturbations through the inherent quantum fluctuations of a field, it is important to determine, from a very general perspective, the observability of the features at short distances compared with $1/H$ through their imprint on this primordial spectrum.  While this imprint is not expected to be observed in the most recent experiments \cite{wmap}, future observations of the microwave background \cite{planck} and of the large scale structure over large volumes of the observed universe \cite{ska} should be able to extract the spectrum of primordial perturbations to a far better precision.

This article has focused on showing the consistency of this effective theory \cite{greens,dSgreens} in terms of its effect on the background geometry.  An effective theory, broadly speaking, corresponds to a perturbative expansion based on the smallness of the ratio between the energy scales being experimentally tested and the scale at which new phenomena can appear---here $H/M$.  In the case of the effective initial state, we include the effects of these new phenomena through a general set of short-distance structures in the state.  The consistency of this expansion requires a suitable prescription for absorbing the divergences that occur in any calculation that sums over all scales and is thus sensitive to these new short-distance structures.  

An important instance of this behavior occurs when we determine the effect of a quantum field on a classical gravitational background.  The energy-momentum tensor provides the source for gravity and in taking the expectation value of this tensor we are implicitly summing over all momentum scales of the field configuration.  Here we have found that in evaluating this sum, we encounter new divergences related to the short-distance structure included in the state.  These divergences are confined to the initial time at which the state is defined and are proportional to purely geometric objects, derivatives of the scale factor in the case of a Robertson-Walker space-time.  These divergences are removed by adding a three-dimensional initial boundary action which contains the appropriate set of geometric counterterms.  This boundary renormalization thereby renders a finite set of gravitational equations of motion.

A second aspect to the consistency of an effective theory is that even after the new corrections from the short-distance features of the state have been rendered finite through renormalization, they should also be small when compared with analogous long-distance effects.  In an inflationary setting this requirement means that the contribution from the short-distance structures of the state to the expectation value of the energy-momentum tensor, which is often called the {\it backreaction\/} \cite{schalm2,backreact,simeone}, should be negligible compared with the vacuum energy of the field that sustains the inflationary expansion.  We found in Eq.~(\ref{backreact}) that the leading trans-Planckian effect is already suppressed by a factor of $(H^2/M_{\rm pl}^2)(H/M)$ and that subleading effects are suppressed by further powers of $H/M$.

The purpose of this work has been to lay a consistent foundation for predicting how the details of physics above the Hubble scale during inflation can affect observations of the universe at large scales or at early times.  The first stage \cite{greens,dSgreens} was to show that the loop corrections for an effective initial state are renormalizable and small in an interacting field theory and here we have shown the renormalizability of the energy-momentum tensor through an appropriate modification of the background, specifically along the space-like surface where the state is defined.  The approach we have followed is fairly standard \cite{books}, reducing both sides of the gravitational equations of motion first to classical quantities.  However, while such an approach is conceptually simpler, the new structures in the state are inherently quantum objects, being the interference between the initial state and subsequent sources.  Therefore a more complete analysis of the energy-momentum renormalization would treat the gravitational component as a quantum field theory as well, thus avoiding the need to impose constraints such as Eq.~(\ref{Nconstraint}) since the time-evolution of expectation value of the gravitational field contains an extra time integral.  This analysis will be done in \cite{next}.  

Once we have established the renormalizability of a theory with an arbitrary effective initial state, it then becomes possible to add the effect of the leading generic trans-Planckian signal to fits of the observed cosmic background radiation spectrum.  The prediction for a particular model of the origin of this signal---whether as a deformed symmetry \cite{smolin}, as a modified uncertainty relation \cite{gary} or as a composite inflaton \cite{cliffnote}, among many further possibilities---is extracted by evaluating the coefficient of the leading moment of the short-distance structure in that model.

\begin{acknowledgments}

\noindent
This work was supported in part by DOE grant No.~DE-FG03-91-ER40682 and the National Science Foundation grant No.~PHY02-44801.  

\end{acknowledgments}

\vskip36truept

\appendix
\section{The adiabatic expansion of the density and pressure}
\label{adiabatic4}

To calculate the divergences associated with the leading trans-Planckian effective state, we need an expansion of the integrands in the general expressions of the pressure and energy density to a sufficient order in the adiabatic expansion that captures the explicit structure of the terms decaying no faster than $k^{-4}$.  In this appendix we show the adiabatic expansion to this order.  To do so, we still only require retaining the terms up to the first adiabatic correction, 
\begin{eqnarray}
\Omega_k &\!\!\!=\!\!\!& 
\omega_k - {1\over 2} {1\over\omega_k} {a^{\prime\prime}\over a} 
- {1\over 4} {a^2m^2\over\omega_k^3} 
\biggl[ {a^{\prime\prime}\over a} + \left( {a'\over a} \right)^2 \biggr] 
\nonumber \\
&&
+\ {1\over 8} {1\over\omega_k^3} 
\biggl[ 
{a^{\prime\prime\prime\prime}\over a} 
- 2 {a^{\prime\prime\prime}\over a} {a'\over a} 
- 2 \left( {a^{\prime\prime}\over a} \right)^2 
+ 2 {a^{\prime\prime}\over a} \left( {a'\over a} \right)^2 
\biggr] 
\nonumber \\
&&
+ {\cal O} \biggl( {1\over\omega_k^5} \biggr) 
\label{adiabat4}
\end{eqnarray}
Inserting these terms into the expressions for $\rho_{\rm surf}(\eta)$ and $p_{\rm surf}(\eta)$ in Eqs.~(\ref{rhosurfO})--(\ref{psurfO}) we find that 
\begin{widetext}
\begin{eqnarray}
\rho_{\rm surf}(\eta)
&=& {1\over 4} {1\over a^4} \int {d^3\vec k\over (2\pi)^3}\, 
f_k^* e^{-2i\int_{\eta_0}^\eta d\eta'\, \Omega_k(\eta')} 
\nonumber \\
&&\quad
\times\biggl\{ 
{1\over\omega_k} \biggl[ 
{a^{\prime\prime}\over a} + \left( {a^{\prime}\over a} \right)^2 
\biggr]
+ {1\over 2} {a^2m^2\over\omega_k^3} \biggl[ 
{a^{\prime\prime}\over a} + 3 \left( {a^{\prime}\over a} \right)^2 
\biggr]
- {1\over 4} {1\over\omega_k^3} \biggl[ 
{a^{\prime\prime\prime\prime}\over a} 
- 3 \left( {a^{\prime\prime}\over a} \right)^2 
- 2 {a^{\prime\prime}\over a} \left( {a'\over a} \right)^2 
\biggr]
\nonumber \\
&&\qquad
+ 2i {a'\over a} 
+ i {a^2m^2\over\omega_k^2} {a'\over a} 
- {i\over 2} {1\over\omega_k^2} \biggl[ 
{a^{\prime\prime\prime}\over a} 
- {a^{\prime\prime}\over a} {a'\over a}
\biggr]
- {i\over 4} {a^2m^2\over\omega_k^4} \biggl[ 
{a^{\prime\prime\prime}\over a} 
- {a^{\prime\prime}\over a} {a'\over a}
\biggr]
\nonumber \\
&&\qquad
+ {i\over 8} {1\over\omega_k^4} \biggl[
{a^{\prime\prime\prime\prime\prime}\over a} 
- 3 {a^{\prime\prime\prime\prime}\over a} {a'\over a}
- 8 {a^{\prime\prime\prime}\over a} {a^{\prime\prime}\over a} 
+ 6 {a^{\prime\prime\prime}\over a} \left( {a'\over a} \right)^2 
+ 10 \left( {a^{\prime\prime}\over a} \right)^2 {a'\over a} 
- 6 {a^{\prime\prime}\over a} \left( {a'\over a} \right)^3 
\biggr]
+ {\cal O} \biggl( {1\over\omega_k^5} \biggr) 
\biggr\}
\label{rhosurfA4}
\end{eqnarray}
and
\begin{eqnarray}
p_{\rm surf}(\eta)
&=& {1\over 4} {1\over a^4} \int {d^3\vec k\over (2\pi)^3}\, 
f_k^* e^{-2i\int_{\eta_0}^\eta d\eta'\, \Omega_k(\eta')} 
\nonumber \\
&&\quad
\times\biggl\{ 
- {4\over 3} \omega_k 
- {2\over 3} {a^2m^2\over\omega_k}
+ {1\over 3} {1\over\omega_k} \biggl[ 
{a^{\prime\prime}\over a} + 3 \left( {a^{\prime}\over a} \right)^2 
\biggr]
- {1\over 6} {a^2m^2\over\omega_k^3} \biggl[ 
{a^{\prime\prime}\over a} - 7 \left( {a^{\prime}\over a} \right)^2 
\biggr]
\nonumber \\
&&\qquad
- {1\over 12} {1\over\omega_k^3} \biggl[ 
{a^{\prime\prime\prime\prime}\over a} 
+ 4 {a^{\prime\prime\prime}\over a} {a'\over a} 
- \Bigl( {a^{\prime\prime}\over a} \Bigr)^2 
- 10 {a^{\prime\prime}\over a} \left( {a'\over a} \right)^2 
\biggr]
\nonumber \\
&&\qquad
+ 2i {a'\over a} 
+ i {a^2m^2\over\omega_k^2} {a'\over a} 
- {i\over 2} {1\over\omega_k^2} \biggl[ 
{a^{\prime\prime\prime}\over a} 
- {a^{\prime\prime}\over a} {a'\over a}
\biggr]
- {i\over 4} {a^2m^2\over\omega_k^4} \biggl[ 
{a^{\prime\prime\prime}\over a} 
- {a^{\prime\prime}\over a} {a'\over a}
\biggr]
\nonumber \\
&&\qquad
+ {i\over 8} {1\over\omega_k^4} \biggl[
{a^{\prime\prime\prime\prime\prime}\over a} 
- 3 {a^{\prime\prime\prime\prime}\over a} {a'\over a}
- 8 {a^{\prime\prime\prime}\over a} {a^{\prime\prime}\over a} 
+ 6 {a^{\prime\prime\prime}\over a} \left( {a'\over a} \right)^2 
+ 10 \left( {a^{\prime\prime}\over a} \right)^2 {a'\over a} 
- 6 {a^{\prime\prime}\over a} \left( {a'\over a} \right)^3 
\biggr] 
+ {\cal O} \biggl( {1\over\omega_k^5} \biggr) 
\biggr\} . 
\label{psurfA4}
\end{eqnarray}
\end{widetext}


\begin{thebibliography}{99}


\bibitem{greens}
H.~Collins and R.~Holman,
Phys.\ Rev.\ D {\bf 71}, 085009 (2005) [hep-th/0501158].

\bibitem{dSgreens}
H.~Collins and R.~Holman,
hep-th/0507081.

\bibitem{schalm}
K.~Schalm, G.~Shiu and J.~P.~van der Schaar,
JHEP {\bf 0404}, 076 (2004) [hep-th/0401164]; 
%
B.~Greene, K.~Schalm, J.~P.~van der Schaar and G.~Shiu,
eConf {\bf C041213}, 0001 (2004) [astro-ph/0503458].

\bibitem{schalm2}
B.~R.~Greene, K.~Schalm, G.~Shiu and J.~P.~van der Schaar,
JCAP {\bf 0502}, 001 (2005) [hep-th/0411217];
%
K.~Schalm, G.~Shiu and J.~P.~van der Schaar,
AIP Conf.\ Proc.\  {\bf 743}, 362 (2005) [hep-th/0412288].

\bibitem{ekp}
R.~Easther, W.~H.~Kinney and H.~Peiris,
JCAP {\bf 0505}, 009 (2005) [astro-ph/0412613]; 
%
R.~Easther, W.~H.~Kinney and H.~Peiris,
astro-ph/0505426.

\bibitem{emil}
P.~R.~Anderson, C.~Molina-Paris and E.~Mottola,
Phys.\ Rev.\ D {\bf 72}, 043515 (2005) [hep-th/0504134].

\bibitem{eft}
H.~Georgi, ``Weak Interactions And Modern Particle Theory,'' (Benjamin/Cummings, Menlo Park, California, 1984); 
%
H.~Georgi, ``Effective field theory,''
Ann.\ Rev.\ Nucl.\ Part.\ Sci.\  {\bf 43}, 209 (1993); 
%
I.~Z.~Rothstein,
``TASI lectures on effective field theories,''
hep-ph/0308266.

\bibitem{kaloper}
N.~Kaloper, M.~Kleban, A.~E.~Lawrence and S.~Shenker,
Phys.\ Rev.\ D {\bf 66}, 123510 (2002); 
%
N.~Kaloper, M.~Kleban, A.~Lawrence, S.~Shenker and L.~Susskind,
JHEP {\bf 0211}, 037 (2002).

\bibitem{cliff}
C.~P.~Burgess, J.~M.~Cline, F.~Lemieux and R.~Holman,
JHEP {\bf 0302}, 048 (2003) [hep-th/0210233]; 
%
C.~P.~Burgess, J.~M.~Cline and R.~Holman,
JCAP {\bf 0310}, 004 (2003) [hep-th/0306079]; 
%
C.~P.~Burgess, J.~Cline, F.~Lemieux and R.~Holman,
astro-ph/0306236.

\bibitem{brandenberger}
J.~Martin and R.~H.~Brandenberger,
Phys.\ Rev.\ D {\bf 63}, 123501 (2001) [hep-th/0005209]; 
%
R.~H.~Brandenberger and J.~Martin,
Mod.\ Phys.\ Lett.\ A {\bf 16}, 999 (2001) [astro-ph/0005432].

\bibitem{gary}
R.~Easther, B.~R.~Greene, W.~H.~Kinney and G.~Shiu,
Phys.\ Rev.\ D {\bf 64}, 103502 (2001); 
%
R.~Easther, B.~R.~Greene, W.~H.~Kinney and G.~Shiu,
Phys.\ Rev.\ D {\bf 67}, 063508 (2003); 
%
G.~Shiu and I.~Wasserman,
Phys.\ Lett.\ B {\bf 536}, 1 (2002); 
%
R.~Easther, B.~R.~Greene, W.~H.~Kinney and G.~Shiu,
Phys.\ Rev.\ D {\bf 66}, 023518 (2002).

\bibitem{transplanck}
J.~C.~Niemeyer,
Phys.\ Rev.\ D {\bf 63}, 123502 (2001); 
%
A.~Kempf,
Phys.\ Rev.\ D {\bf 63}, 083514 (2001); 
%
J.~C.~Niemeyer and R.~Parentani,
Phys.\ Rev.\ D {\bf 64}, 101301 (2001); 
%
A.~Kempf and J.~C.~Niemeyer,
Phys.\ Rev.\ D {\bf 64}, 103501 (2001); 
%
A.~A.~Starobinsky,
Pisma Zh.\ Eksp.\ Teor.\ Fiz.\  {\bf 73}, 415 (2001)
[JETP Lett.\  {\bf 73}, 371 (2001)]; 
%
L.~Hui and W.~H.~Kinney,
Phys.\ Rev.\ D {\bf 65}, 103507 (2002); 
%
S.~Shankaranarayanan,
Class.\ Quant.\ Grav.\  {\bf 20}, 75 (2003); 
%
S.~F.~Hassan and M.~S.~Sloth,
Nucl.\ Phys.\ B {\bf 674}, 434 (2003); 
%
K.~Goldstein and D.~A.~Lowe,
Phys.\ Rev.\ D {\bf 67}, 063502 (2003); 
%
V.~Bozza, M.~Giovannini and G.~Veneziano,
JCAP {\bf 0305}, 001 (2003); 
%
G.~L.~Alberghi, R.~Casadio and A.~Tronconi,
Phys.\ Lett.\ B {\bf 579}, 1 (2004); 
%
J.~Martin and R.~Brandenberger,
Phys.\ Rev.\ D {\bf 68}, 063513 (2003); 
%
U.~H.~Danielsson,
Phys.\ Rev.\ D {\bf 66}, 023511 (2002); 
%
U.~H.~Danielsson,
JHEP {\bf 0207}, 040 (2002); 
%
R.~H.~Brandenberger and J.~Martin,
Int.\ J.\ Mod.\ Phys.\ A {\bf 17}, 3663 (2002).  

\bibitem{fate}
H.~Collins, R.~Holman and M.~R.~Martin,
Phys.\ Rev.\ D {\bf 68}, 124012 (2003) [hep-th/0306028]; 
%
H.~Collins and M.~R.~Martin,
Phys.\ Rev.\ D {\bf 70}, 084021 (2004) [hep-ph/0309265]; 
%
H.~Collins,
hep-th/0312144.

\bibitem{textbooks}
A.~Linde, {\it Particle Physics and Inflationary Cosmology\/} (Harwood Academic, Chur, Switzerland, 1990) [hep-th/0503203]; 
E.~W.~Kolb and M.~S.~Turner, {\it The Early Universe\/} (Addison-Wesley, Reading, MA, 1990); 
A.~Liddle and D.~Lyth, {\it Cosmological Inflation and Large-Scale Structure\/} (Cambridge University Press, Cambridge, England,
2000).

\bibitem{dodelson}
S.~Dodelson, {\it Modern Cosmology\/} (Academic Press, San Diego, CA, 2003).  

\bibitem{schwinger}
J.~S.~Schwinger,
J.\ Math.\ Phys.\  {\bf 2}, 407 (1961); 
%
L.~V.~Keldysh,
Zh.\ Eksp.\ Teor.\ Fiz.\  {\bf 47}, 1515 (1964)
[Sov.\ Phys.\ JETP {\bf 20}, 1018 (1965)]; 
%
K.~T.~Mahanthappa, Phys.\ Rev. {\bf 126}, 329 (1962); P.~M.~Bakshi and K.~T.~Mahanthappa, J.\ Math.\ Phys.\ {\bf 4}, 12 (1963).

\bibitem{next}
H.~Collins and R.~Holman, 
hep-th/0609002.

\bibitem{wmap}
H.~V.~Peiris {\it et al.},
Astrophys.\ J.\ Suppl.\  {\bf 148}, 213 (2003) [astro-ph/0302225]; 
%
D.~N.~Spergel {\it et al.},
astro-ph/0603449.

\bibitem{bunch}
T.~S.~Bunch and P.~C.~Davies,
Proc.\ Roy.\ Soc.\ Lond.\ A {\bf 360}, 117 (1978).

\bibitem{einhorn}
M.~B.~Einhorn and F.~Larsen, 
Phys.\ Rev.\ D {\bf 67}, 024001 (2003) [hep-th/0209159].

\bibitem{alpha}
N.~A.~Chernikov and E.~A.~Tagirov,
Annales Poincare Phys.\ Theor.\ {\bf A9}, 109 (1968); 
E.~A.~Tagirov,
Annals Phys.\  {\bf 76}, 561 (1973); 
%
E.~Mottola,
Phys.\ Rev.\ D {\bf 31}, 754 (1985); 
%
B.~Allen,
Phys.\ Rev.\ D {\bf 32}, 3136 (1985).

\bibitem{books}
N.~D.~Birrell and P.~C.~W.~Davies, {\it Quantum Fields in Curved Space\/} (Cambridge University Press, Cambridge, UK, 1982); 
%
R.~M.~Wald, {\it General Relativity\/} (The University of Chicago Press, Chicago, IL, 1984); 
%
S.~M.~Fulling, {\it Aspects of Quantum Field Theory in Curved Spacetime\/} (Cambridge University Press, Cambridge, UK, 1989).

\bibitem{adiabatic}
L.~Parker and S.~A.~Fulling,
Phys.\ Rev.\ D {\bf 9}, 341 (1974); 
%
S.~A.~Fulling and L.~Parker,
Annals Phys.\  {\bf 87}, 176 (1974); 
%
S.~A.~Fulling, L.~Parker and B.~L.~Hu,
Phys.\ Rev.\ D {\bf 10}, 3905 (1974); 
%
T.~S.~Bunch,
J.\ Phys.\ A {\bf 13}, 1297 (1980); 
%
P.~R.~Anderson and L.~Parker,
Phys.\ Rev.\ D {\bf 36}, 2963 (1987); 
%
N.~D.~Birrell,
Proc.\ Roy.\ Soc.\ Lond.\ {\bf 361}, 513 (1978).

\bibitem{pointsplit}
S.~M.~Christensen,
Phys.\ Rev.\ D {\bf 14}, 2490 (1976).

\bibitem{gh}
G.~W.~Gibbons and S.~W.~Hawking,
Phys.\ Rev.\ D {\bf 15}, 2752 (1977).

\bibitem{wald}
See, for example, Appendix E of 
R.~M.~Wald, {\it General Relativity\/} (The University of Chicago Press, Chicago, IL, 1984).

\bibitem{he}
S.~W.~Hawking and G.~F.~R.~Ellis, {\it The large-scale structure of space-time\/} (Cambridge University Press, Cambridge, UK, 1973). 

\bibitem{planck}
J.~A.~Tauber on behalf of ESA and the Planck Scientific Collaboration, 
Adv.\ Space Res.\ {\bf 34}, 491 (2004).

\bibitem{ska}
Two such examples include the Square Kilometre Array and the Cosmic Inflation Probe:  respectively, 
S.~Rawlings, F.~B.~Abdalla, S.~L.~Bridle, C.~A.~Blake, C.~M.~Baugh, L.~J.~Greenhill and J.~M.~van der Hulst,
New Astron.\ Rev.\  {\bf 48}, 1013 (2004) [astro-ph/0409479] 
and  
G.~J.~Melnick, G.~G.~Fazio, V.~Tolls, D.~T.~Jaffe, K.~Gebhardt, V.~Bromm, E.~Komatsu, and R.~A.~Woodruff,
Am.\ Astron.\ Soc.\ Meeting {\bf 205} (2004) 10006.

\bibitem{backreact}
M.~Porrati,
Phys.\ Lett.\ B {\bf 596}, 306 (2004) [hep-th/0402038].
%
M.~Porrati,
hep-th/0409210.
%
F.~Nitti, M.~Porrati and J.~W.~Rombouts,
hep-th/0503247.

\bibitem{simeone}
D.~Lopez Nacir, F.~D.~Mazzitelli and C.~Simeone,
Phys.\ Rev.\ D {\bf 72}, 124013 (2005) [gr-qc/0511046].

\bibitem{smolin}
G.~Amelino-Camelia, L.~Smolin and A.~Starodubtsev,
Class.\ Quant.\ Grav.\  {\bf 21}, 3095 (2004) [hep-th/0306134]; 
%
L.~Smolin,
hep-th/0501091; 
%
L.~Smolin,
hep-th/0605052.

\bibitem{cliffnote} 
We would like to thank Cliff Burgess for suggesting this example to us.

\end{thebibliography}
\end{document}